\documentclass[twocolumn]{aastex62}
\usepackage{natbib}
\usepackage{amsmath}
\usepackage{textcomp}
\usepackage{booktabs}
\usepackage{bookmark}
\usepackage{color}
\usepackage{soul}
\usepackage{url}
\usepackage{enumitem}

\hypersetup{linkcolor=blue, citecolor=blue}
\shorttitle{Examining \ensuremath{\lambda_{7.7}} with PAHdb}
\shortauthors{M.~J. Shannon \& C. Boersma}

\begin{document}
\newcommand{\mt}{\textmu m~}
\newcommand{\m}{\textmu m}
\newcommand{\HII}{\mbox{H\,\textsc{ii}}}%
\newcommand{\msun}{M\ensuremath{_{\odot}}}
\newcommand{\Wpeak}{\ensuremath{\lambda_{7.7}}}
\newcommand{\teff}{\ensuremath{T_{\text{eff}}}}
\newcommand{\hae}{HAeBe~}
\newcommand{\nc}{$N_{\rm c}$}
\newcommand{\ang}{\mbox{\normalfont\AA}}

\title{Examining the class B-to-A shift of the 7.7~\mt PAH band with the NASA Ames PAH IR Spectroscopic Database}

\author[0000-0001-5681-5151]{Matthew J. Shannon}
\altaffiliation{NASA Postdoctoral Program Fellow}
\affiliation{Universities Space Research Association, Columbia, MD}
\affiliation{NASA Ames Research Center, MS 245-6, Moffett Field, CA 94035-1000}

\author[0000-0002-4836-217X]{Christiaan Boersma}
\affiliation{NASA Ames Research Center, MS 245-6, Moffett Field, CA 94035-1000}

\correspondingauthor{Matthew J. Shannon}
\email{matthew.j.shannon@nasa.gov}

\begin{abstract}
We present insights into the behavior of the astronomical 7.7~\mt polycyclic aromatic hydrocarbon (PAH) emission complex as gleaned from analyzing synthesized spectra, utilizing the data and tools from the NASA Ames PAH IR Spectroscopic Database. We specifically study the influence of PAH size, charge, aliphatic content and nitrogen substitution on the profile and peak position of the 7.7~\mt feature (\Wpeak). The 7.7~\mt band is known to vary significantly from object-to-object in astronomical observations, but the origin of these variations remains highly speculative. Our results indicate that PAH size can accommodate the largest shift in \Wpeak~($\simeq0.4$~\m), where relatively small PAHs are consistent with class~A spectra (\nc $\leq 60$) while large PAHs are consistent with red/very red class~B spectra. Aliphatic PAHs, of which our sample only contains a few, can produce redshifts typically around $0.15$~\m; changes in ionization fraction, depending on the species, produce shifts up to $0.1$~\m; and nitrogen substitution has no effect on \Wpeak. Within the limits of our study, the class B{\textrightarrow}A transition is best explained with a changing PAH size distribution, with a relatively minor role assigned to aliphatic content and varying charge states. The resulting astronomical picture is that the photochemical evolution of PAHs moving from shielded class C/B environments into exposed ISM-like class A environments may be intrinsically different from the reverse class~A{\textrightarrow}B transition of interstellar PAHs being incorporated into newly-forming star systems.
\end{abstract}



\section{Introduction}
\label{sec:intro}

A prominent family of infrared (IR)-emitting molecules in space are the polycyclic aromatic hydrocarbons (PAHs), which are composed of interlocking hexagonal carbon cycles and typically a peripheral coating of hydrogen atoms. PAHs in the interstellar medium (ISM) and circumstellar environments are thought to contain on average $N_{\rm c}=50-100$ carbon atoms \citep[e.g.,][]{allamandola1989}. Spectroscopically, they can dominate the IR appearance of objects, particularly since their (mid-IR) 3-20~\mt emission bands are often observed at very high contrast relative to the continuum \citep{allamandola1989}. PAHs are found in many (distinct) environments, including reflection nebulae, planetary nebulae, dusty pre-main sequence objects, the general ISM and carbon-rich evolved stars. For a thorough review of interstellar PAHs, see \citet{tielens2008}.

Not only are the prominent PAH emission features highly variable in both absolute and relative strength, the profiles of the individual bands are known to vary as well---sometimes significantly. The most striking in this regard is the 7.7~\mt PAH emission complex: its peak position (\Wpeak~hereafter) is observed to shift from 7.6 (in mostly ISM-type sources) to 7.8 (in mostly circumstellar-type environments) to beyond 8~\mt (as far as $8.5$~\m, in dusty objects). A systematic PAH profile classification scheme was developed by \citet{peeters2002} and \citet{vandiedenhoven2004}. Under this scheme, when the 7.7~\mt complex peaks near 7.6~\mt it is class~A; emission peaking near 7.8~\mt is class~B; and emission that peaks beyond 8~\mt is class~C. \citet{matsuura2014} expanded this classification scheme to include class~D, defined by a broad 7.7~\mt complex that is relatively flat between 7.6--7.8~\mt with a suppressed 8.6~\mt band.

\begin{figure*}
\centering
\includegraphics[width=0.7\linewidth]{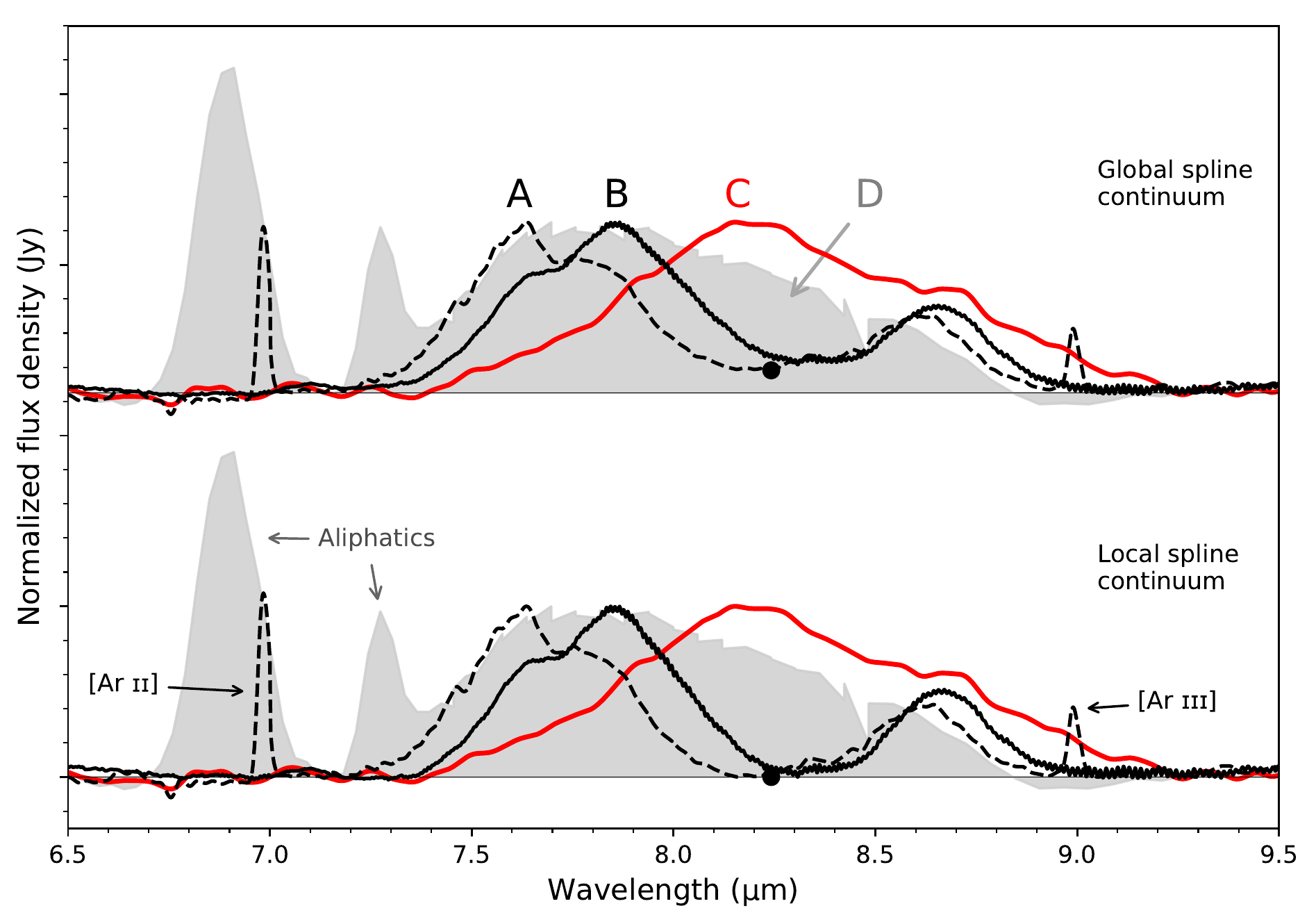}
\caption{The four classes (A, B, C, D---shaded region) of the 7.7~\mt PAH complex \citep{peeters2002,vandiedenhoven2004,matsuura2014}. The classes are broadly associated with emission peaking near 7.6 (A), 7.8 (B), beyond 8.0 (C), or a broad feature generically coincident with the class~A and B positions and a missing 8.6 \mt feature (D). Note that a broadband spline continuum has been subtracted from these data. Whether or not an anchor point near 8.2~\mt (black dot) is incorporated into the spline determines whether it is a so-called global spline or local spline, respectively. See Section~\ref{sec:analysis} for details regarding the continuum determination. We utilize three ISO/SWS sources to represent each class's spectrum: IRAS 23133+6050 (class~A), HD~44179 (class~B), IRAS~13416-6243 (class~C); and one Spitzer/IRS source, IRAS~05110-6616 (class~D). The class~C spectrum has been smoothed for clarity. Note that the class~A-C sources are the prototypes for their classes as presented by \citet{peeters2002} and \citet{vandiedenhoven2004}.}
\label{fig:classes}
\end{figure*}

Spectra illustrating each PAH profile class are shown in Fig.~\ref{fig:classes}, utilizing data from the Short Wavelength Spectrometer (SWS; \citealt{degraauw1996}) of the Infrared Space Observatory (ISO; \citealt{kessler1996}) and the Infrared Spectrograph (IRS; \citealt{houck2004}) of the Spitzer Space Telescope \citep{werner2004}. A broadband spline continuum has been subtracted from these spectra (see Section~\ref{sec:analysis} for details). A spline with an anchor point near 8.2~\mt represents a ``local'' continuum, whereas one without represents the ``global'' continuum \citep[cf.\ ][]{peeters2012}. This distinction only applies to the class~A and B spectra as the class~C and D spectra are broad and individual subcomponents are difficult to isolate/identify.

Relationships observed between different parameters contain clues about the origin(s) of the class variations: a particularly prominent example is the fact that \Wpeak~is correlated with effective stellar temperature (\teff) of the irradiating source in many objects \citep{sloan2007}. As \teff~decreases, \Wpeak~shifts toward longer wavelengths, which holds for many sources as long as \teff~$\lesssim 12000$~K. Several studies have now confirmed this relationship (e.g., \citealt{boersma2008,keller2008,smolders2010}).

However, the question of \textit{why} there is a correlation between \Wpeak~and \teff~for many objects remains an open problem. Suggestions include the destruction of aliphatic-rich material \citep{sloan2007} or the evolution of an active chemical balance \citep{boersma2008}. Herbig Ae/Be (\hae) stars are of particular interest for this correlation because they have a mostly limited range of \teff~yet span a large range in PAH class, from 7.6 (class A) to as high as 8.4~\mt (class C). There appears to be a distinction between isolated \hae stars and non-isolated \hae stars (i.e., those with surrounding nebulosity). Understanding the difference from one \hae system to the next (and their corresponding changes in \Wpeak) could be key to understanding the origin of the PAH class transitions. 

In this paper we utilize the NASA Ames PAH IR Spectroscopic Database\footnote{\url{www.astrochemistry.org/pahdb/}} (PAHdb; \citealt{boersma2014_amesdb,bauschlicher2018}; Mattioda et al., in progress) to study the profile and peak position of the 7.7~\mt complex for different PAH (sub-)populations. PAHdb is a collection of more than 3,000 theoretically computed spectra and 75 experimentally measured PAH spectra, plus a comprehensive suite of analytical tools and methods. We construct PAH sub-populations from PAHdb by choosing species according to their molecular properties---i.e., size, structure, charge state and composition. We overlay quantities derived from the synthesized spectra on figures showing their observational equivalent, specifically \Wpeak~v.s.~\teff~and \Wpeak~v.s.~the 6.2/11.2~\mt PAH band strength ratio, for direct comparison. We aim to characterize the chemical and physical changes in the PAH populations as the PAH class changes, and place these results in the context of the \Wpeak~vs.~\teff~relationship and the life-cycle of aromatic material through the different phases of stellar evolution. 

We construct synthetic PAH spectra in Section~\ref{sec:pahdb} and examine their \Wpeak~behavior in Section~\ref{sec:results1}. These lead into a discussion of the astronomical context and its relevance in Section~\ref{sec:discussion}, including a summary of the evolution of aromatic material through the different stages of the star- and planet-forming process. We finish with a brief summary of our results and conclusions in Section~\ref{sec:conclusions}.


\section{\texorpdfstring{\Wpeak~}{lambda77} in synthesized PAH spectra}
\label{sec:pahdb}

The NASA Ames PAH IR Spectroscopic Database contains large libraries of theoretically-calculated and experimentally-measured PAH spectra for a variety of PAH sizes, structures, ionization states and atomic (heterogeneous) substitutions. We make use of version 3.00 of the library of computed spectra of PAHdb \citep{bauschlicher2018} and select molecular subsets to study the behavior of \Wpeak~as a function of different molecular parameters. For all reported measurements we require, for the purpose of consistency, that the synthetic spectra we create (and analyze) look \emph{qualitatively} similar to astronomical PAH spectra. That is, the major bands at 6.2, 7.7 and 11.2 \mt should be clearly visible, as well as their broad underlying plateaus.

\subsection{Assumptions and parameters}
\label{sec:pahdb_assumptions}

We construct synthetic PAH spectra from PAHdb in the following manner:

\begin{enumerate}[noitemsep]
    \item Take the collection of ``astronomical'' PAHs from \citet{bauschlicher2018} and
    \item identify the subset that satisfies a chosen criterion (e.g., $N_{\rm c} \geq 90$) with the restriction that both its singly cationic and neutral spectra must be available, then
    \item assume a size distribution,
    \item assume an excitation energy, 
    \item apply a PAH emission model, and lastly,
    \item assume a PAH ionization fraction.
\end{enumerate}

\begin{figure}
\centering
\includegraphics[width=1\linewidth]{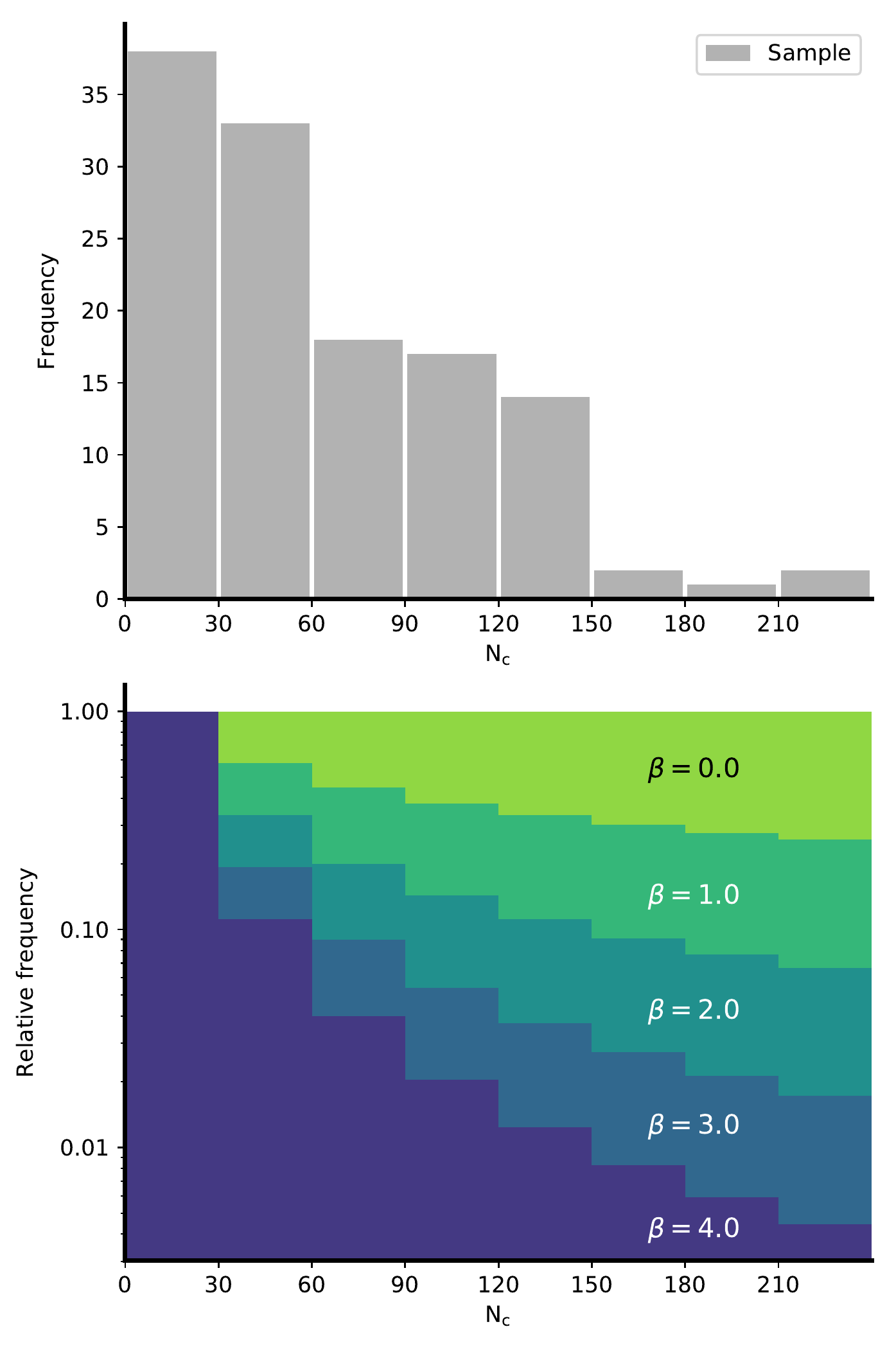}
\caption{\textit{Top:} Histogram depicting the intrinsic size distribution of the 125 pure PAHs in our sample. Species have been placed into bins of size $\Delta N_{\rm c} = 30$ (i.e., $N_{\rm c} \leq 30$, $30 < N_{\rm c} \leq 60$, etc.). \textit{Bottom:} The effect of assuming a power-law size distribution, governed by power-law index $\beta$. For $\beta=0.0$, there is no additional scaling as a function of PAH size; when increasing $\beta$, small PAHs are given additional weight. See Section~\ref{sec:pahdb_assumptions} for details.}
\label{fig:pahdb_histogram}
\end{figure}

\begin{figure}
\centering
\includegraphics[width=1\linewidth]{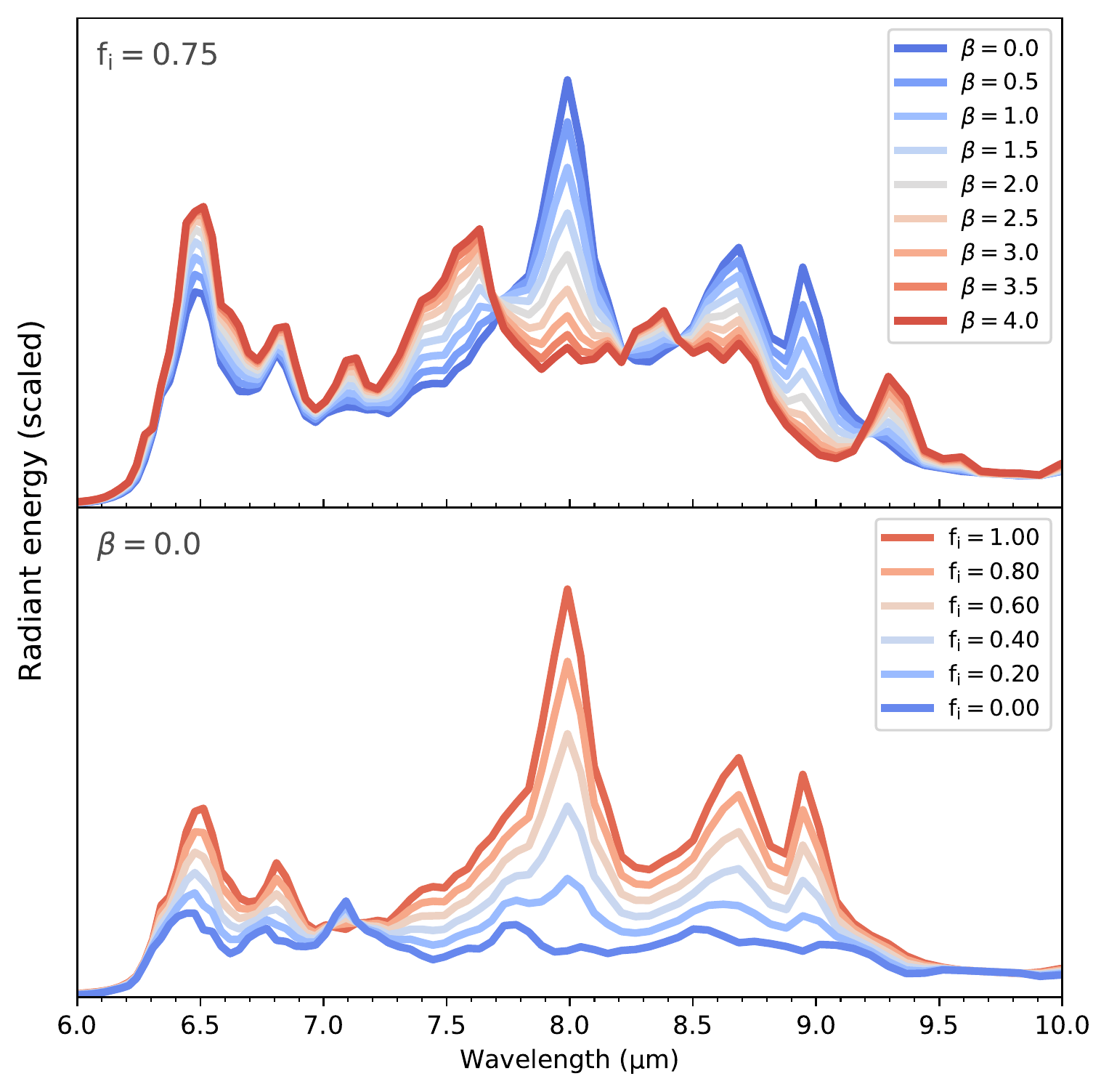}
\caption{Impact of parameter choice on the synthesized PAH spectra (shown here for the pure PAHs in our sample). \textit{Top:} the power-law index $\beta$ is varied between 0.0--4.0 at a fixed ionization fraction of $f_i=0.75$. Higher values of $\beta$ emphasize the contribution from smaller-sized PAHs (cf.\ Fig.~\ref{fig:pahdb_histogram}). \textit{Bottom:} the ionization fraction ($f_i$) is varied while holding the power-law index fixed at $\beta=0.0$. To make reliable measurements of \Wpeak, we require $f_i \geq 0.30$, as the neutral spectrum is typically devoid of a well-defined 7.7~\mt band. For presentation purposes the spectra in each panel are normalized to the spectrum with the maximum peak radiant energy. See Section~\ref{sec:pahdb_assumptions} for details.}
\label{fig:pahdb_fi}
\end{figure}

The astronomical PAH size distribution is often considered to be that of the dust grain size distribution extended into the molecular domain \citep{tielensbook}.
The grain size distribution is represented by the power-law $\frac{\mathrm{d}n}{\mathrm{d}a} \propto a^{-\beta}$, where $n$ is the number density of grains, $a$ is the radius and $\beta$ is the power-law index, typically assumed to be 3.5 for interstellar dust grains (the MRN distribution; \citealt{mathis1977}). Note that the MRN size distribution was defined for species considerably larger than PAH molecules (i.e., $a\gtrsim 50\ \ang$ versus $a\simeq10\ \ang$, respectively). We vary $\beta$ between 0.0--4.0 to determine its influence on the synthesized PAH spectra. We compute an effective radius $a_\text{eff}$ for use in the MRN power-law by computing $a_\text{eff} = (A/ \pi)^{0.5}$, where $A$ is the PAH surface area determined through simply counting rings; $n_{\rm rings}\times A_{\rm single\ ring}$. A histogram of the distribution of PAH sizes is presented in Fig.~\ref{fig:pahdb_histogram} for several choices of $\beta$, which captures the \emph{relative} change in the PAH size distribution. It should be noted that there are a (relatively) limited number of PAH species in PAHdb with more than 50 carbon atoms that meet \emph{all} of the criteria set out above. Example spectra demonstrating the effect of varying $\beta$ are presented in the upper panel of Fig.~\ref{fig:pahdb_fi}. 
We have averaged the spectral contribution from species with the same $a_{\rm eff}$ to prevent double-weighting species with greater representation in PAHdb than other species. In other words, each PAH size  is represented by the average spectrum from all PAHs of that size. Subsequently, each bin is represented by the average spectrum from the spectra in that bin.

In the emission model the PAHs are each excited by a single 5~eV photon and the entire emission cascade is taken into account when calculating the emerging spectrum; see \\citet{boersma2013} for specifics. The choice of 5~eV appears somewhat arbitrary, but as will be explored later in the paper (Sections~\ref{sec:analysis} and \ref{subsec:astro}), the \emph{shape} of the 7.7 \mt complex is insensitive to the choice of photon energy.

A redshift of 15~cm$^{-1}$ is adopted to mimic some anharmonic effects, a common choice found in the literature (see \citealt{bauschlicher2009}, their Section~3, for a brief discussion). We compute spectra using both Lorentzian and Gaussian line profiles, which makes a minor to negligible difference in the resulting spectra, except for systematically introducing a large broadband continuum component in the Lorentzian case. We have chosen to display the results utilizing the Lorentzian line profile with a full-width at half-maximum (FWHM) of 15~$\text{cm}^{-1}$. Some of the limitations associated with the above choices are addressed in Section~\ref{subsec:limitations}.

We can synthesize spectra at any desired ionization fraction ($\text{f}_i \equiv n_{\text{PAH}^{+}} / (n_{\text{PAH}^{0}} + n_{\text{PAH}^{+}})$); see Fig.~\ref{fig:pahdb_fi} (lower panel) for some examples. At low ionization fractions there is minimal emission in the 6--9~\mt region, generally making the determination of \Wpeak~unreliable. In our analysis, we therefore limit ourselves to ionization fractions of 30\% or greater, i.e., $f_i \geq 0.30$.

In total, our sample consists of 125 `pure' PAHs---i.e., those with no aliphatic sidegroups and no heteroatom substitutions--for which both the neutral and singly charged cation's spectra are available. We also include a sample of 16 superhydrogenated PAHs and 25 PAHs with aliphatic sidegroups. PAHs with heteroatom substitutions of nitrogen (PANHs) are also considered. They total 33 molecules and are all small at $N_{\rm c} \leq 23$. In total over 2,000 spectra were synthesized.

\subsection{Methods}
\label{sec:analysis}

The distinct \emph{astronomical} mid-IR PAH emission features reside on broad underlying plateaus, on top of emission from, for example, warm dust, small heated grains and starlight. Several methods have been developed to separate the PAH features from these ``continuum'' components. One commonly used method in the literature is to distinguish the broad underlying emission from the PAH features by fitting a spline through a series of fixed anchor points. (e.g., \citealt{vankerckhoven2000, hony2001, peeters2002, galliano2008b, boersma2014, peeters2017, shannon2018}). Other methods include fitting Drude profiles (e.g., PAHFIT; \citealt{smith2007}) or Lorentzian profiles \citealt{boulanger1998, galliano2008b}).
As a consequence, these methods assign portions of the underlying plateau and broad band continuum emission to the overlapping broad wings characteristic of these profiles. On occasion, individual bands are decomposed with Gaussian profiles to probe spectral substructure \citep[e.g.,][]{shannon2016, peeters2017, stock2017}, including the 7.7 \mt complex, but high-resolution observations are generally better suited for this approach.

Although different methods lead to different implied band intensities, \emph{astronomical} PAH band strength ratios and conclusions derived therefrom are largely insensitive to the chosen method (e.g., \citealt{galliano2008b,peeters2017}).

Here, we treat our synthesized spectra broadly as ``astronomical'' spectra and adopt the spline method to isolate the distinct PAH features from any broader underlying plateau and continuum components. This implies that we consider part of the plateau and continuum emission to originate from PAHs in genuine astronomical spectra. After determining the shape of the underlying broadband emission, we tested four methods for measuring \Wpeak. The simplest approach is to identify the wavelength at which the feature peaks. A slightly more sophisticated approach is to identify the barycenter \citep[cf.\ ][]{sloan2007}, i.e., the wavelength at which the area under the band is halved, also known as the first moment. This approach converges with the results of the first method as the symmetry of the band increases. We also tested a 2-Gaussian and 4-Gaussian decomposition, mimicking the methods of \citet{sloan2007} and \citet{peeters2017}, respectively. It is noted that the 7.6/7.8 \mt band ratio is sometimes used as a PAH class indicator as well \citep[e.g.,][]{boersma2010, boersma2014}. However, this measure does not allow for a direct comparison with the classification scheme set out by \cite{peeters2002} and used by others.

Of the four methods considered, we were able to establish that the choice of method has little effect on the \Wpeak~measurements. The results presented from here onwards are based on the barycenter method. Given the grid on which we model our spectra and the uncertainties involved with drawing the continua, we estimate an accuracy of $\pm 0.01$ \mt on \Wpeak. 

\begin{figure}
\centering
\includegraphics[width=1\linewidth]{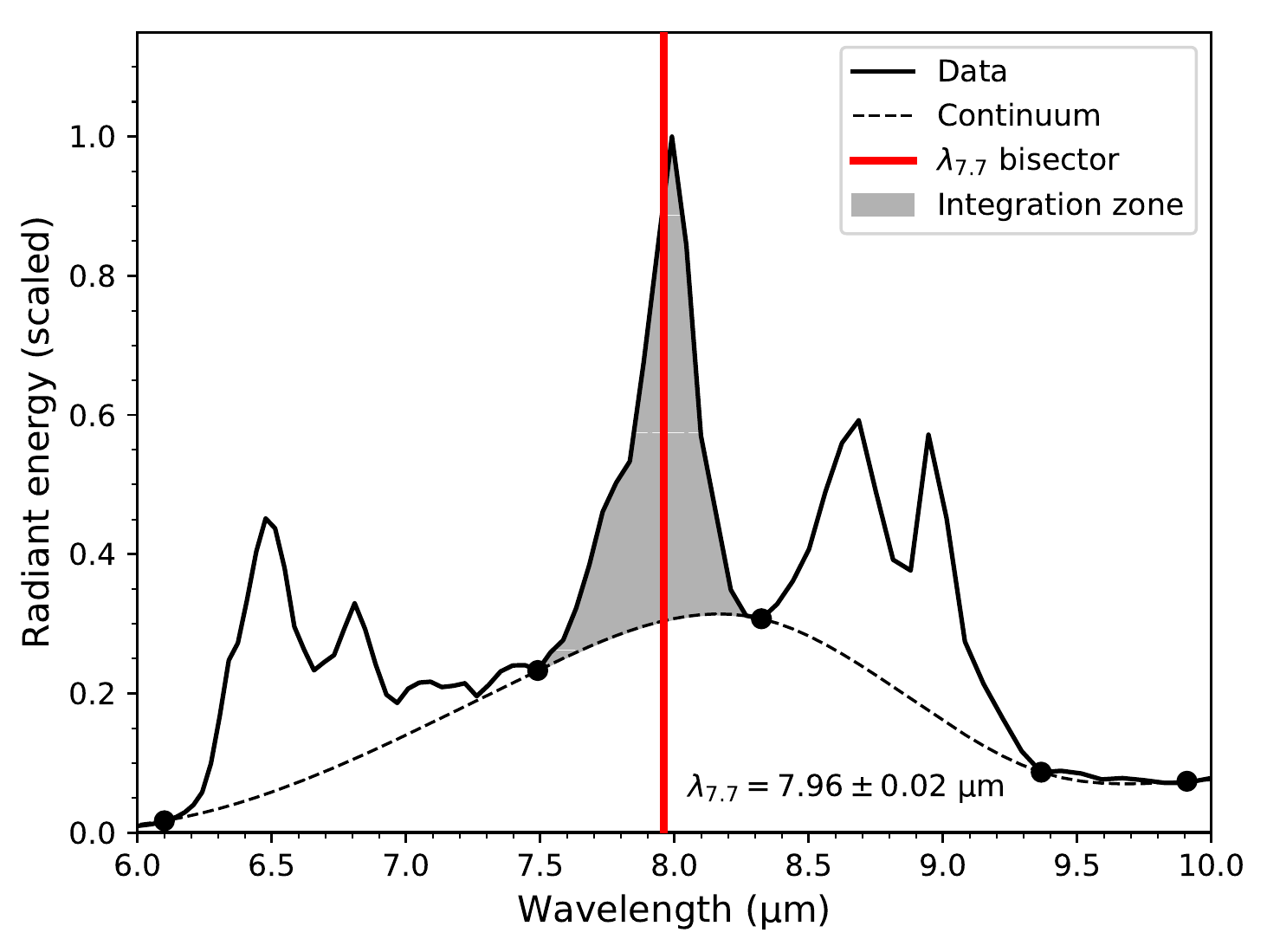}
\caption{Demonstration of the approach used to determine \Wpeak~in our synthesized PAH emission spectra. The broadband emission is fit with a spline (dashed line) defined by a series of anchor points (black dots). The 7.7~\mt PAH band is integrated within the shaded region. The barycenter of the feature (\Wpeak$=7.96\pm0.01$~\mt here) is denoted by the vertical red line. Note that the thickness of the line is about the estimated uncertainty. See Section~\ref{sec:analysis} for details.}
\label{fig:test3}
\end{figure}


\section{Results}
\label{sec:results1}

In Section~\ref{results:size} we first present results in which no inherent size distribution ($\beta$=0.0) is assumed, i.e., all sizes contribute equally. In Section~\ref{sec:results_beta}, we examine the impact of imposing a PAH size distribution (i.e., varying $\beta$). In Sections~\ref{sec:results_aliph}--\ref{results:dehydro} we present \Wpeak~results from considering different PAH subclasses (i.e., the presence of aliphatic material, nitrogenation, and dehydrogenation).

\subsection{The effect of PAH size and charge state on \texorpdfstring{\Wpeak}{lambda77}}
\label{results:size}

\begin{figure*}
\centering
\includegraphics[width=\linewidth]{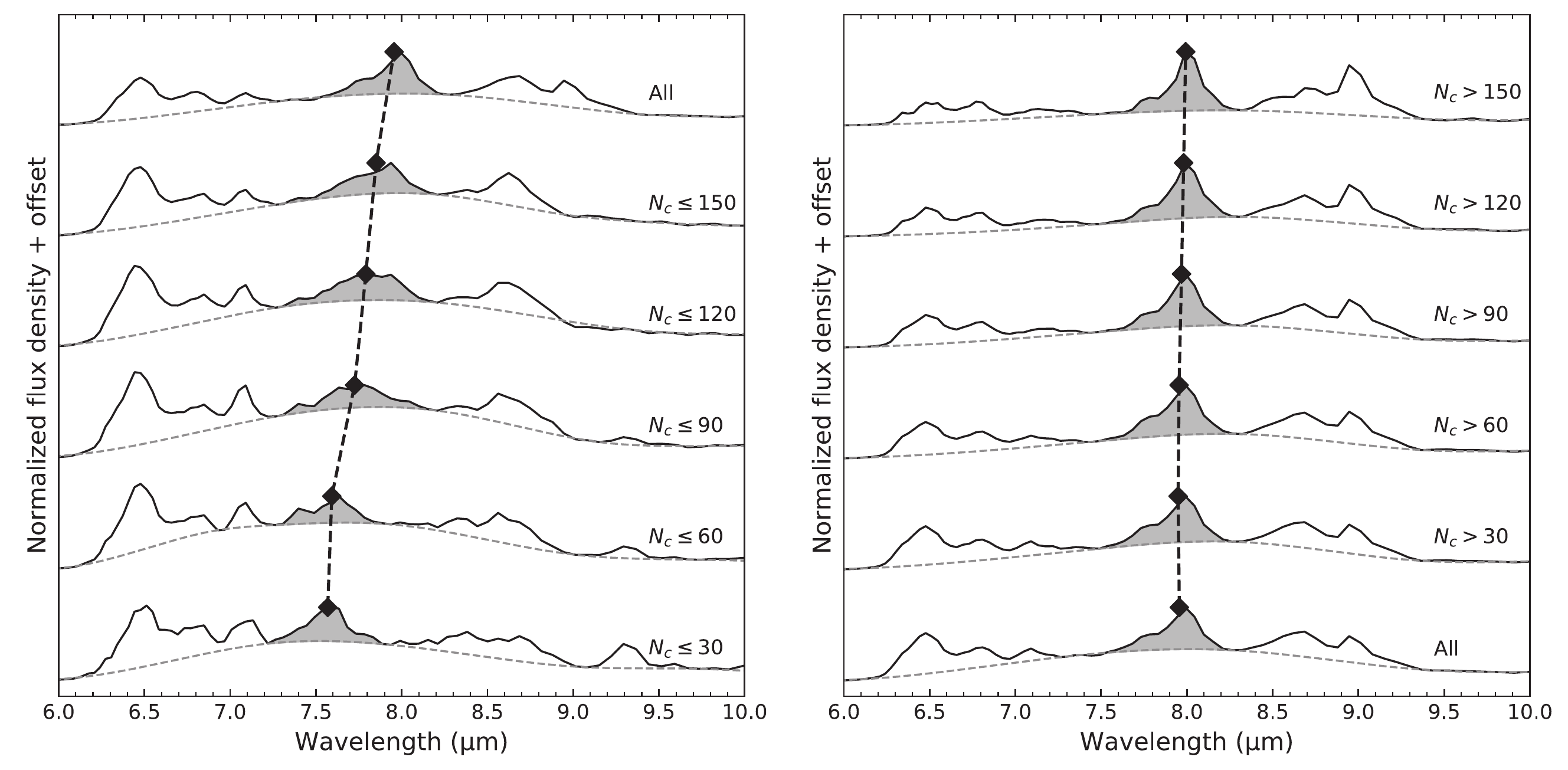}
\caption{Synthesized PAH emission spectra for a fixed ionization fraction of $f_i$=0.40 and a fixed $\beta = 0.0$. \Wpeak~is measured within the shaded region and is denoted by the black diamond symbols. The sample of synthesized spectra is truncated by excluding large species (Case 1, left panel) or small species (Case 2, right panel), as measured by the number of carbon atoms ($N_{\rm c}$). The dashed lines show a spline fit to the broadband emission underlying the more distinct PAH features. A summary of results for \Wpeak~versus ionization fraction for multiple PAH sizes is presented in Fig.~\ref{fig:pahdb_size_vs_fi}. See Section~\ref{results:size} for details.}
\label{fig:pahdb_size}
\end{figure*}

\begin{figure}
\centering
\includegraphics[width=\linewidth]{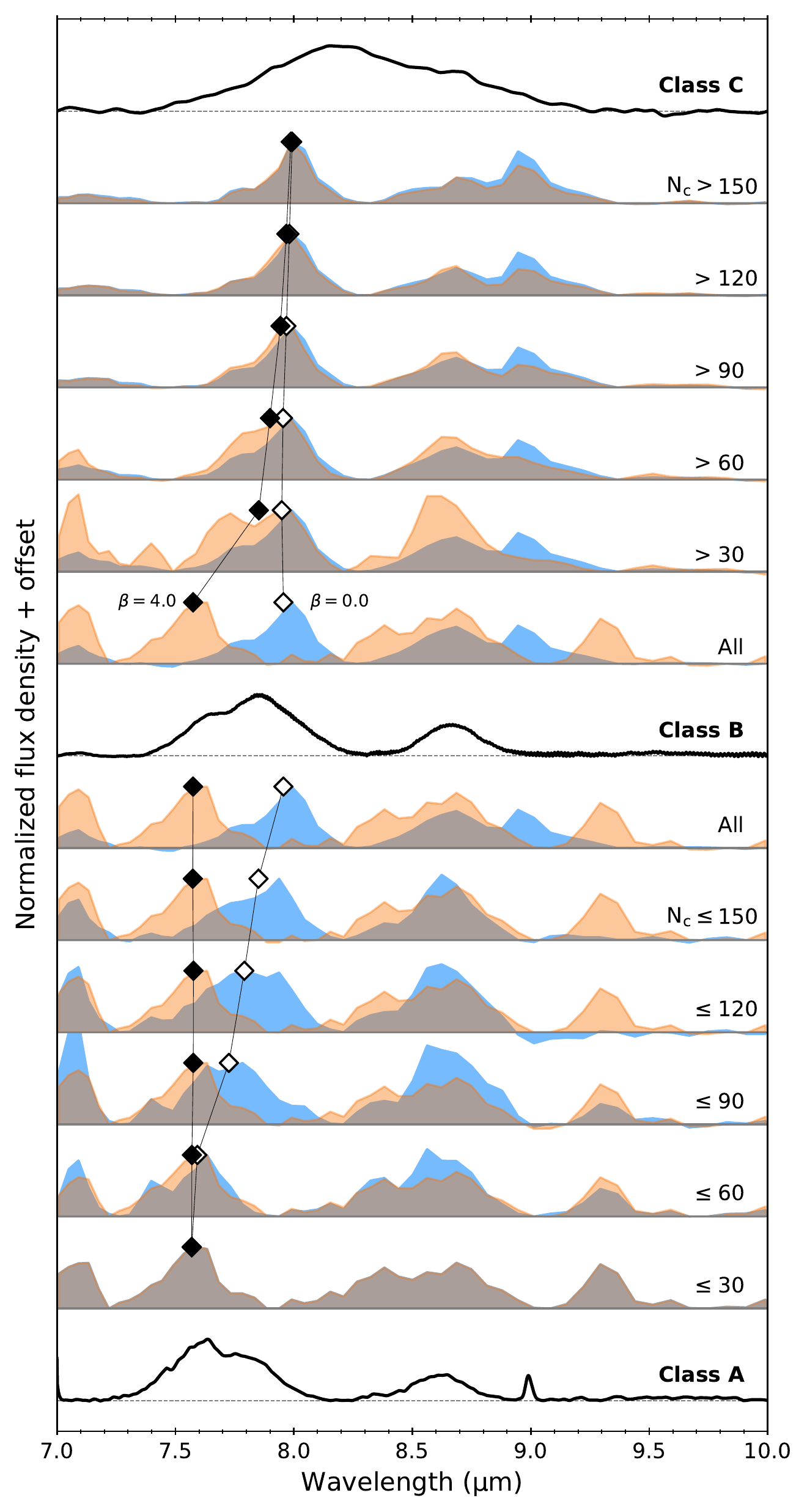}
\caption{Comparison of continuum-subtracted synthesized PAH emission spectra (at a fixed ionization fraction of $f_i=0.40$) to the prototypical astronomical class~A, B and C spectra. The spectra for two values of the power-law index $\beta$ are shown: $\beta=0.0$ (in the blue shading), which allows strong contributions from large PAHs, and $\beta=4.0$ (in the orange shading), which emphasizes the emission from the smaller PAHs. NB the brownish color arises from the overlap of the blue and orange spectra. Note the disparity between class~C and \emph{all} other spectra. See Section~\ref{results:size} for details.}
\label{fig:pahdb_sizesub1}
\end{figure}

We start out by segregating the species in our sample of pure PAHs by number of carbon atoms ($N_{\rm c}$), where $N_{\rm c}$ is used as a proxy for PAH size. PAHs are collected into bins of $\Delta N_{\rm c} = 30$ (i.e., $N_{\rm c} \leq 30$, $30 < N_{\rm c} \leq 60$, etc.). Next, two cases are considered: 1. dropping PAH sizes starting at the large end, and 2. conversely dropping PAH sizes starting at the small end. Figs.~\ref{fig:pahdb_size} \& \ref{fig:pahdb_sizesub1} present the evolution of the synthesized spectra as certain PAH sizes are dropped and at a fixed ionization fraction of $f_{\rm i} = 0.4$. The latter figure shows the results when subtracting the broadband spline continuum, depicted in the former figure, from the spectra and makes a direct comparison with the prototypical spectra of the astronomical PAH classes.

\paragraph{Case~1} The left panel of Fig.~\ref{fig:pahdb_size} and the lower half of Fig.~\ref{fig:pahdb_sizesub1} present the results when dropping PAH sizes starting from the large end. Perusal of the figures shows a monotonic blueshift of \Wpeak~as more-and-more larger PAHs are dropped. \Wpeak~start out as high as 7.95~\mt and generally shows good agreement with the astronomical class~B spectrum. When only considering the smallest PAHs, i.e., $N_{\rm c} \leq 30$ or $N_{\rm c} \leq 60$, \Wpeak~is consistent with a class~A position at 7.5--7.6~\mt but with a stronger 8.6~\mt band relative to the 7.7~\mt feature and blueshifted by $ \sim$0.05~\mt from the nominal class~A position (7.6~\m).

\paragraph{Case~2} The right panel of Fig.~\ref{fig:pahdb_size} and the upper half of Fig.~\ref{fig:pahdb_sizesub1} present the results when dropping PAH sizes starting from the small end. We observe a mild monotonic redshift of \Wpeak~as more-and-more small PAHs are dropped, but not as profound as the blueshifts seen in Case~1. \Wpeak~starts out at $\sim$7.9~\mt and generally shows good agreement with the astronomical class~B spectrum. When only considering the largest PAHs (e.g., those with $N_{\rm c} \geq 120$ or $N_{\rm c} \geq 150$), \Wpeak~shifts into the 7.98--8.00~\mt range and remains consistent with astronomical class~B PAH spectra. At no point is similarity with the class~C profile achieved.

Thus overall, changes in the PAH ionization fraction only introduce a small shift in \Wpeak, generally comparable to changing the large end cut-off by $\Delta N_{\rm c} = 30$. We note that the wavelength shift of \Wpeak~as a function of $N_{\rm c}$ is monotonic when considering bins of $\Delta N_{\rm c} = 30$. This trend holds when examining the dataset with finer granularity (e.g., $\Delta N_{\rm c} = 20$), but at the threshold of $\Delta N_{\rm c} = 10$, one runs into sampling issues as the PAHs in our sample are not equally represented across all $N_{\rm c}$ (see Fig.~\ref{fig:pahdb_histogram}).

\begin{figure*}
\centering
\includegraphics[width=0.85\linewidth]{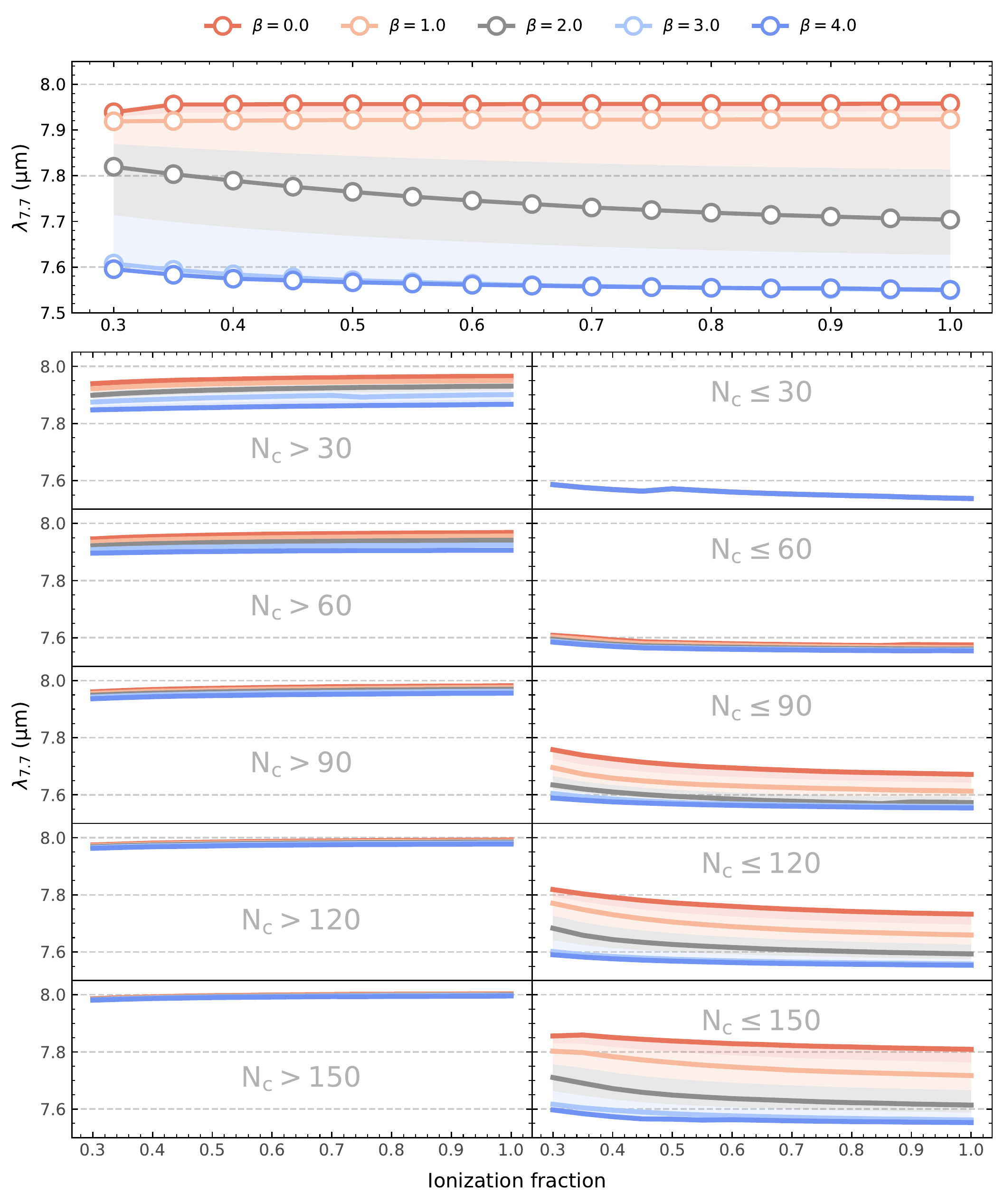}
\caption{The effect of imposing different size ($N_{\rm c}$) and power-law ($\beta$) constraints on the PAH size distribution as a function of ionization fraction ($f_i$) on \Wpeak. \textit{Upper-most panel:} \Wpeak~vs.\ ionization fraction ($f_i$) as a function of $\beta$ for the entire sample of pure PAHs. The shaded zones are to help guide the eye and the thin horizontal lines denote the boundaries of classes A, B and C (7.6, 7.8 and 8.0 \m, respectively). \textit{Left stacked panels:} Small PAHs are selectively clipped from the distribution, resulting in a strong shift of \Wpeak~towards longer wavelengths. \textit{Right stacked panels}: Large PAHs are selectively clipped, further emphasizing the contribution from the smallest PAHs. See Sections~\ref{results:size}~\& \ref{sec:results_beta} for details.}
\label{fig:pahdb_size_vs_fi}
\end{figure*}

\subsection{The effect of \texorpdfstring{$\beta$}{beta} on \texorpdfstring{\Wpeak}{lambda77}}
\label{sec:results_beta}

We now impose a PAH size distribution by varying $\beta$ (see Section~\ref{sec:pahdb_assumptions}). As illustrated in Fig.~\ref{fig:pahdb_histogram}, increasing $\beta$ will quickly emphasize the contribution from small PAHs. This can be observed in Fig.~\ref{fig:pahdb_sizesub1}, where the spectra for $\beta=4.0$ (when considering all the pure PAHs) have a \Wpeak~near 7.55 \m, whereas for $\beta=0.0$ \Wpeak~is near 7.95 \m. By examining different small-size and large-size cut-offs of our sample (cf.\ Section~\ref{results:size}), we verify that PAH size is responsible for this large shift in \Wpeak. Due to the nature of the power-law size distribution, the removal of small PAHs (with non-zero $\beta$ values) leads to significant changes in the synthesized spectra (Fig.~\ref{fig:pahdb_sizesub1}). Conversely, very little happens when large PAHs are removed instead. We note that the strengths of the 8.6 \mt band (relative to the 7.7 \mt band) in the synthesized spectra are higher than observed in astronomical spectra. Qualitatively, the spectra show good agreement with the astronomical class~A and B spectra.

Fig.~\ref{fig:pahdb_size_vs_fi} summarizes the position of \Wpeak~as a function of ionization fraction and $\beta$ for our sample and all its subsets. Naturally, high values of $\beta$ enhance the contribution from small PAHs, which tend to peak near the astronomical class~A position. Thus, higher values of $\beta$ are linked to a blueshifted \Wpeak~peak position, while lower values of $\beta$ allow a greater contribution from large PAHs, which redshift \Wpeak.

\subsection{The effect of aliphatic-bonded material on \texorpdfstring{\Wpeak}{lambda}}
\label{sec:results_aliph}

\begin{figure}
\centering
\includegraphics[width=1\linewidth]{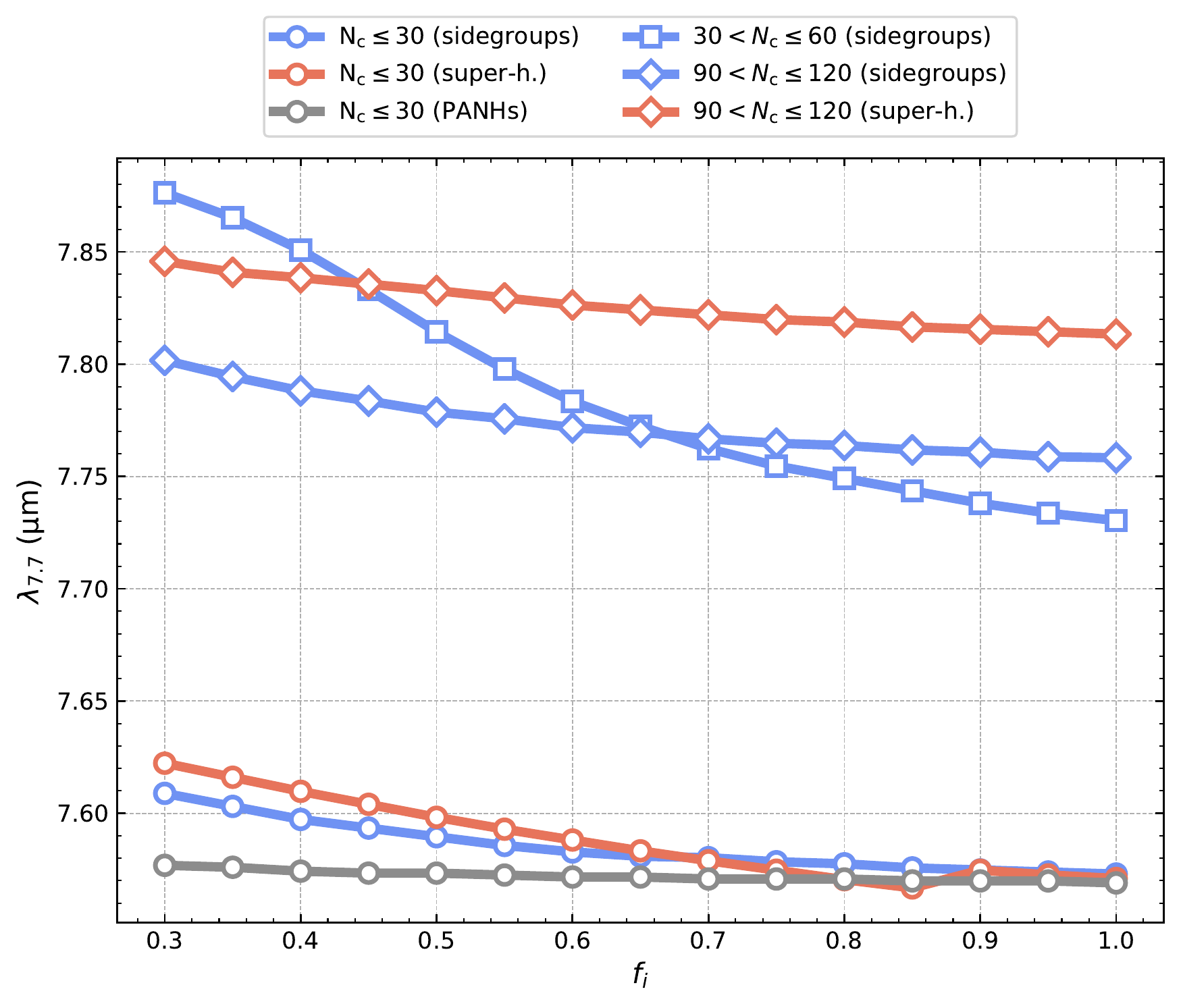}
\caption{Summary of the variations in \Wpeak~measured on synthesized emission spectra of PANHs, superhydrogenated PAHs and PAHs with aliphatic sidegroups, as a function of ionization fraction ($f_i$). See Section~\ref{sec:results_aliph} for details.}
\label{fig:pahdb_size_vs_fi_aliph}
\end{figure}

Our sample contains a group of 16 superhydrogenated and a group of 25 PAHs with aliphatic sidegroups (subject to the restriction that both the neutral and singly ionized spectra should be present). While these are few in number, they are potentially important subsets as variations in aliphatic content was suggested by \citet{sloan2007} to explain the evolution of \Wpeak~with \teff.

Our \Wpeak~results are summarized in Fig.~\ref{fig:pahdb_size_vs_fi_aliph}. For large superhydrogenated PAHs ($90 < N_c \leq 120$), we observe that \Wpeak~varies from 7.82--7.85~\m, where a higher ionization fraction is associated with a blueshifted \Wpeak. The small superhydrogenated PAHs ($N_c \leq 30$) are systematically blueshifted from the large molecules, spanning approximately $7.57-7.62$~\m. Clearly, size is again important in determining \Wpeak.

The synthesized spectra of PAHs with sidegroups also show a strong separation by size: small PAHs with sidegroups ($N_c \leq 30$) peak between 7.57--7.61~\m, while large species ($90 < N_c \leq 120$) peak between 7.76--7.80~\m, depending on ionization fraction. Intermediate-sized PAHs with aliphatic sidegroups ($30 < N_c \leq 60$) are somewhat atypical, spanning 7.73--7.88 \m, with a stronger dependence on ionization than the other subsets. This is likely due to the limited number of PAHs with aliphatic sidegroups in our sample, as it is hard to average-out variations between specific molecules. Nonetheless, PAHs with aliphatic sidegroups generally behave like superhydrogenated PAHs. The astronomical class~A and B positions (roughly 7.6 and 7.8~\m, respectively) are easily reproduced by these species, and red B positions at certain ionization fractions, but the astronomical class~C remains elusive (\Wpeak$\geq 8$~\m). We conclude by stating that variations in PAH size of these molecules can account for a total shift of up to $\sim$0.25-0.30 \m, and ionization fraction a shift of up to $\sim$0.15~\mt (though typically less).

\subsection{The effect of nitrogenation on \texorpdfstring{\Wpeak}{lambda77}}
\label{sec:panhs}

We consider the influence of nitrogen substitution into the PAH skeletal structure on \Wpeak~in Fig.~\ref{fig:pahdb_size_vs_fi_aliph}. Note that these species are all relatively small with less than 24 carbon atoms. We find that these PANHs are almost completely insensitive to changes in ionization fraction, with \Wpeak~being nearly constant at 7.57~\m. As such, the PANHs here solely match the astronomical class~A spectrum.

\subsection{The effect of dehydrogenation on \texorpdfstring{\Wpeak}{lambda77}}
\label{results:dehydro}

We also examined the synthesized spectra (cf.\ Section~\ref{sec:analysis}) of dehydrogenated PAHs \citep[see also][]{mackie2015}. We selected the four fully dehydrogenated PAHs from the library of computed spectra of PAHdb for which both the neutral and singly charged spectra are available (i.e., C$_{\rm 24}$, C$_{\rm 54}$, C$_{\rm 66}$, and C$_{\rm 96}$).

There is significant spectral variability in these spectra, including features that are not seen in the astronomical spectra. As such, reliably quantifying \Wpeak~turned out to be impossible. However, two interesting characteristics were observed in these spectra: 1. As the ionization fraction increases, a broad emission band grows near 8 \mt having a FWHM$\simeq$0.7 \m, and 2. A narrow emission band centered around 8.25 \mt remains unchanged in both position and intensity with changing ionization fraction, i.e., $0\leq f_{i}\leq1$.

The presence of features that are not mirrored by the astronomical observations necessitates a healthy degree of skepticism, but these species have a 7.7 \mt band with a very red class~B appearance.


\section{Discussion}
\label{sec:discussion}

We discuss our results by first providing the astronomical context in Section~\ref{subsec:astro}, which is followed by a discussion of the molecular factors that affect \Wpeak~in Section~\ref{sec:discuss_factors}. PAH excitation/destruction and its effect on \Wpeak~is discussed in Section~\ref{sec:discuss_size}, after which the limitations of our study are considered (Section~\ref{subsec:limitations}). Lastly, in Section~\ref{subsec:picture} we paint the picture of aromatic evolution through the different stages of the star- and planet formation process.

\subsection{Astronomical context}
\label{subsec:astro}

Based on a sample of astronomical spectra spanning a variety of objects types, \citet{sloan2007} were able to show that there is a correlation between \Wpeak~and the effective temperature (\teff) of the irradiating star. This correlation is said to reflect the chemical evolution of the PAH populations due to UV photo-processing (from class~C{\textrightarrow}B{\textrightarrow}A). In other words, class~C environments would have a predominance of unprocessed material, and over time, this material would evolve to look like typical ISM class~A PAH spectra. From the analysis of spatial-spectral maps, \citet{bregman2005} showed that in reflection nebulae, more UV exposure shifts the 7.7 band towards shorter wavelengths. \citet{sloan2007} propose the molecular changes that drive the spectral evolution is the destruction of aliphatic-bonded material when transitioning from benign class~C to harsher class~A environments.

Several other studies have confirmed the correlation between \Wpeak~and \teff~and have posited alternative explanations for the PAH class transition beyond the aliphatic hypothesis \citep[e.g.,][]{keller2008, boersma2008, cerrigone2009, acke2010, smolders2010, sloan2014, maaskant2014, seok2017, blasberger2017, raman2017, lazareff2017}. These alternative explanations invoke variable PAH sizes and/or size distributions \citep{bauschlicher2009}; changes in ionization fraction \citep{bauschlicher2008, bauschlicher2009}; or perhaps heteroatom substitution, which is considered important by some to explain the class~A 6.2~\mt band position \citep{peeters2002, hudgins2005}.

\begin{figure*}
\centering
\includegraphics[height=2.6in]{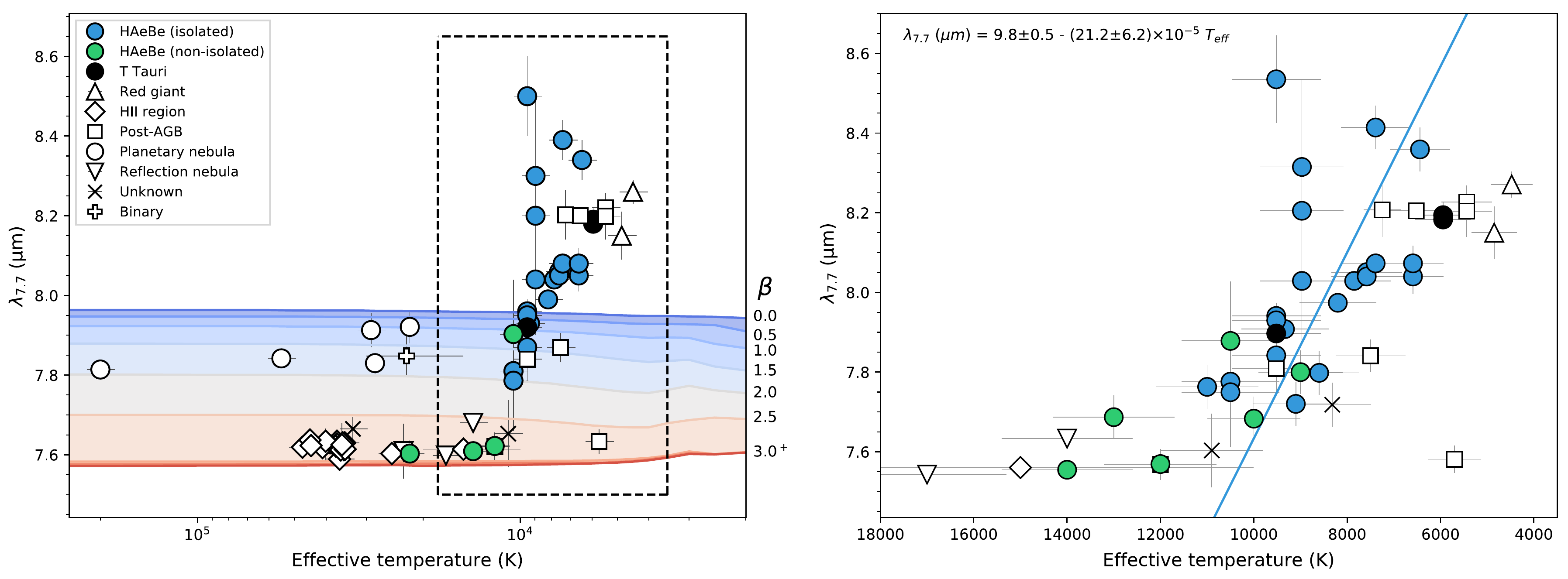}
\caption{\textit{Left:} \Wpeak~vs.~\teff~for our sample astronomical observations from the literature (Table~\ref{tab:sources}). They are over-plotted on measurements of \Wpeak~on synthesized PAH emission spectra. These spectra were synthesized using our sample of pure PAHs at a fixed ionization fraction of $f_i=0.70$ (cf.\ Section~\ref{results:size}) and varying $\beta$ values (indicated by the red-blue shading). Note that the results for $\beta=3.0, 3.5$ and $4.0$ overlap. See also Fig.~\ref{fig:pahdb_size_vs_fi}. \textit{Right:} Zoom-in on the outlined (dashed) region in the left panel, where \Wpeak~increases significantly over a narrow range in \teff, encompassing mainly \hae~stars. The best-fit line (blue) to the data and its equation are also shown. Note that this panel is plotted on a linear scale, unlike the left panel which uses a logarithmic scale. See Section~\ref{subsec:astro} for details.}
\label{fig:77vsteff}
\end{figure*}

Fig.~\ref{fig:77vsteff} reproduces the correlation between \Wpeak~and \teff~for a sample of sources we collected from the literature (e.g., \citealt{peeters2002, sloan2007, keller2008}; see Table.~\ref{tab:sources} in the appendix for a full list of references). This collection contains data on a variety of pre- and post-main sequence objects, including a considerable number of \hae stars. These data originate from spectroscopic observations acquired by Spitzer/IRS and ISO/SWS. In the figure we identify a horizontal spread of sources near \Wpeak$\simeq$7.6~\mt from approximately \teff$\simeq$47,000--14,000~K; a ``knee'' around 14000~K; and a generally linear trend of increasing \Wpeak~as \teff~continues to decrease. A few planetary nebulae do not match the overall trend due to their high \teff. The numerous sources at \Wpeak$\simeq$7.6~\mt predominately consist of \HII-regions, with a few reflection nebulae, a few sources of questionable object type\footnote{Notably, MWC~922, which is thought to be a B[e] emission line star, IRAS~18576+0341 and S~106 (IRS4).}, and a few non-isolated \hae stars. Isolated \hae stars generally exhibit peak positions between $\sim$7.8--8.4~\mt and follow a linear trend (see Fig.~\ref{fig:77vsteff}, right panel, for an expanded view). The isolated and non-isolated \hae systems appear to overlap in the class~B{\textrightarrow}A transition region (7.6--7.8~\m). Red giant and post-AGB stars also fall upon the linear trend line at low \teff, apart from one post-AGB object (IRAS 16279-4757).

Fig.~\ref{fig:77vsteff} also shows the results from varying the PAH size distribution via $\beta$ on \Wpeak. Here we considered all pure PAHs in our sample and a fixed ionization fraction of $f_i=0.70$, though the results are quite insensitive to ionization state (cf. Fig.~\ref{fig:pahdb_size_vs_fi}). The effective stellar temperature (\teff) is used to describe a blackbody, which is subsequently multiplied with the PAH-specific absorption cross-section as described by \cite{draine2007} to derive an average excitation energy. From thereon, the synthetic spectra are constructed as outlined in Section~\ref{sec:pahdb_assumptions}.

The class~A sources with \Wpeak$\simeq$7.6~\mt generally match a population dominated by small or relatively small PAHs, which arises for $\beta \gtrsim2.5$. This is somewhat counter-intuitive, as large PAHs are generally considered more stable than their smaller counterparts and are thus expected to be better represented in the harsher environments of these ISM-type objects. The class~B ($\sim$7.8~\m) and very red class~B sources (approaching 8.0~\m), which are generally circumstellar-type environments, reflect a moderate to strong contribution from larger PAHs. 

PAHs are commonly considered to be an extension of the grain size distribution into the molecular domain, which reflects the intimate relationship between PAHs and the larger grains \citep[see e.g.,][]{tielensbook}. The fact that smaller PAHs contribute near 7.6~\mt and larger PAHs near 7.8~\mt is intrinsic to the PAHs considered in this study and has previously been pointed out by, e.g., \citet{bauschlicher2008, bauschlicher2009}. Thus even if the PAH size distribution would not follow a power-law relationship, this intrinsic dichotomy would always result in a redder \Wpeak~when the population is comprised of a greater number of large PAHs relative to small PAHs. Note that varying the ionization fraction $f_{i}$ does not move away from the fact that redder \Wpeak~positions are associated with larger PAHs.

\subsubsection{Herbig Ae/Be stars}

As can be clearly seen from Fig.~\ref{fig:77vsteff}, \hae stars occupy a special place in the \Wpeak~vs. \teff~relationship. \hae stars have a limited range in \teff, but span a wide range in \Wpeak~and provide a smooth transition between the class~C and B regimes. \hae systems are the pre-main sequence precursors of intermediate-mass stars (2--8~\msun) of spectral types B, A and F \citep{herbig1960}. Such systems house a circumstellar disk and are in some cases still enshrouded by their natal envelope. \citet{boersma2008} suggest that the smooth class~B{\textrightarrow}A transition could be due to a varying degree of class~B and A-like material in the telescope beam (e.g., from the disk and cloud, respectively). However, while the class~C{\textrightarrow}A transition for the conversion of post-AGB star material into ISM material suits an evolutionary interpretation (cf.\ Section~\ref{subsec:astro}), this is not the case when transitioning `back' to class~B circumstellar disk material.

Protoplanetary disks are important environments in the view of chemical complexity, as molecules are released from icy grains and formed, destroyed and reformed. The disk itself provides shielding from the protostar, and thus a generally subdued environment for chemical reactions. Simple molecules such as CO$_2$, NH$_3$ and HCN have been observed in \hae systems proving active, ongoing chemistry \citep{henning2013}. Therefore, ongoing active chemistry in the disk provides a compelling explanation for what drives the PAHs, in an evolutionary sense, `back' from class~A{\textrightarrow}B. 

\subsubsection{The 6.2/11.2\texorpdfstring{~\mt}{micron} PAH band strength ratio and \texorpdfstring{\Wpeak}{lambda77}}
\label{subsubsec:proxy}

The 6.2/11.2~\mt PAH band strength ratio is a qualitative measure of the PAH ionization fraction \citep[e.g.,][]{allamandola1999}. We measure the 6.2 and 11.2~\mt PAH band strengths after subtracting the broadband emission from our synthesized pure PAH emission spectra (cf.\ Section~\ref{results:size}).

\begin{figure}
\centering
\includegraphics[width=1\linewidth]{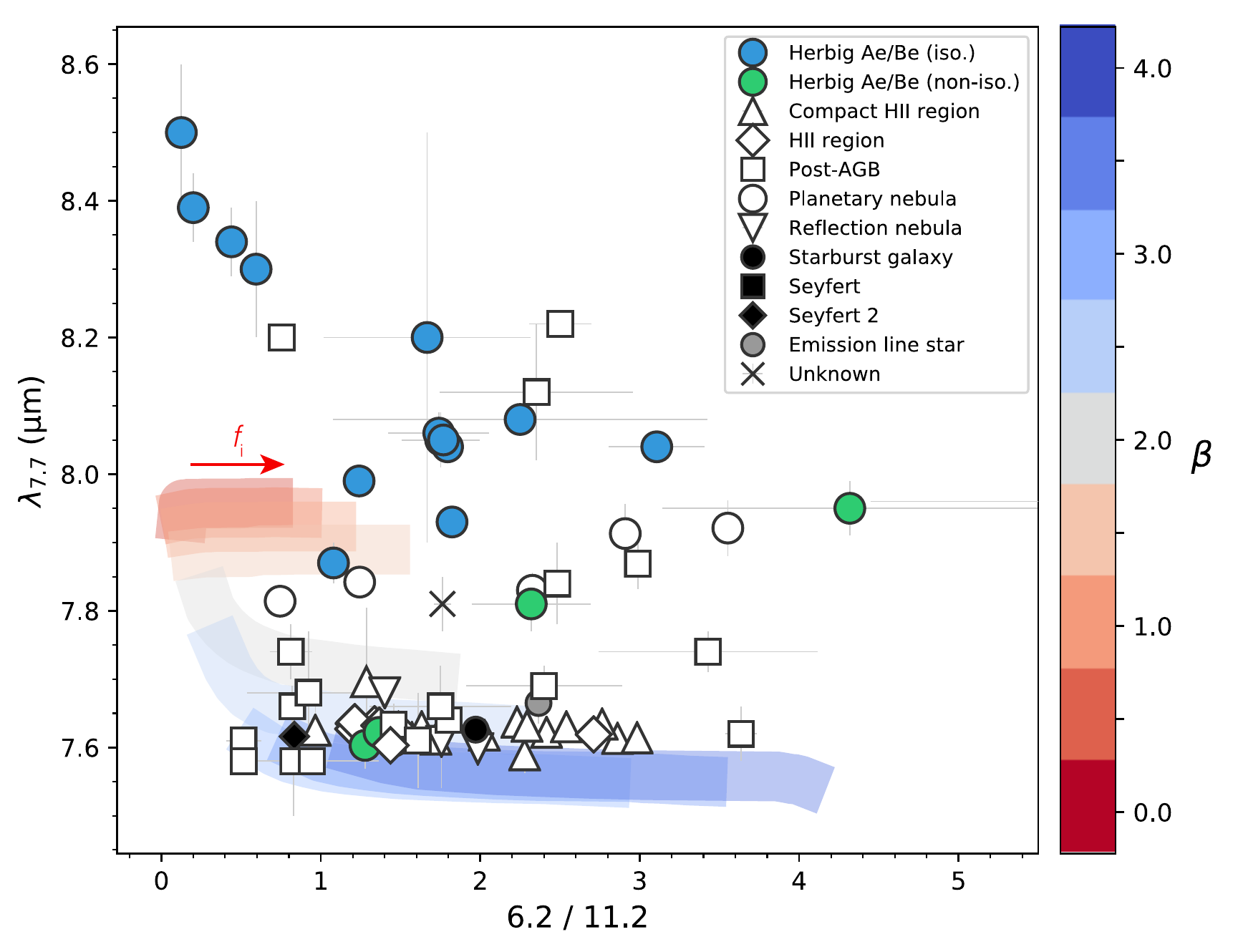}
\caption{\Wpeak~vs.\ the 6.2/11.2~\mt PAH band strength ratio as determined for our sample of astronomical observations (Table~\ref{tab:sources}). We also show the results from our synthesized pure PAH spectra as a function of $\beta$ (red-blue shading) and ionization fraction ($f_{\rm i})$. The (mostly) horizontal span for a given $\beta$ reflects the variation in the degree of ionization. See Section~\ref{subsubsec:proxy} for details.}
\label{fig:ratio_62112}
\end{figure}

In Fig.~\ref{fig:ratio_62112} we take our entire observational dataset, plot the 6.2/11.2~\mt PAH band strength ratio vs.\ \Wpeak, and compare it to measurements of the same quantities on our synthesized spectra. We vary $\beta$ and the ionization fraction ($f_{\rm i}$). The figure shows a natural congregation of sources near the 7.6~\mt (class~A) position, generally from \HII-regions (compact and otherwise), some post-AGB stars, reflection nebulae and a few included galaxies. PAHs with values of $\beta\geq2.5$ are consistent with this cluster of data points. In general, increasing PAH size is associated with a shift towards the upper left (generally higher \Wpeak~and lower 6.2/11.2). As expected, the 6.2/11.2 ratio is well correlated with the ionization fraction ($f_i$), as variations in $f_i$ are linked to a mostly horizontal spread of 6.2/11.2 for any particular $\beta$.

Compared to Fig.~\ref{fig:77vsteff}, there is noticeable scatter for the data points that do not fall on the \Wpeak=7.6 \mt line, especially for the \hae stars. This suggests that the factors that determine \Wpeak~in \hae stars is different from those in other objects. By varying $\beta$, pure PAHs can capture the B{\textrightarrow}A transition, but not from class~C{\textrightarrow}B. The class~B and C sources themselves do not show a clear trend between \Wpeak~and the 6.2/11.2 \mt PAH band strength ratio, though we note that the highest \Wpeak~values are only present at low 6.2/11.2 ratios.

\subsection{Molecular factors affecting \texorpdfstring{\Wpeak}{lambda77}}
\label{sec:discuss_factors}

\paragraph{Size.} We find that overall, and within any given PAH subclass, size is the most important driver of \Wpeak. Varying PAH size produces the largest shifts, generally from 7.6 to almost 8.0~\m. \citet{bauschlicher2008, bauschlicher2009} find that the 7.7~\mt band is generally explained by a mixture of cation and anion contributions from large PAHs (near 7.8~\m) and small PAHs (near 7.6~\m). In other words, the class~A and B differences would be driven by different mean PAH sizes \citep[small and large, respectively, as demarcated by $N_{\rm c}$ = 100;][]{bauschlicher2009}. As far as class~C is concerned, the maximum \Wpeak~we measure ($\sim$7.98~\m) are from synthesized spectra with a strong contribution from very large PAHs ($N_{\rm c} \geq 150$). Per \citet{ricca2012}, PAHs of this size produce 6--9~\mt emission that is complex and generally inconsistent with the overall simplicity of the astronomical spectra. However, PAHs of these sizes could still be present at low abundance when mixed in with PAHs of smaller size (see, e.g., Fig.~\ref{fig:pahdb_sizesub1}). Database-fitting of astronomical spectra also show that small PAHs are important for emission near 7.6~\mt and large PAHs for emission near 7.8~\mt \citep[e.g.,][]{cami2011, boersma2013}. 

In contrast, \citet{seok2017} analyzed spectra of 14 T~Tauri stars and 55 \hae stars with the astro-PAH model of \citet{li2001} and \citet{draine2007}, finding that decreasing PAH sizes are associated with a redshift of the 7.7~\mt band. This finding might well be an effect of the semi-empirical nature of the astro-PAH model they used. Directly relating measurements of PAH size, using the 3.3/11.2~\mt PAH band strength ratio, with \Wpeak~should elucidate matters.

\paragraph{Aliphatic content.} Aliphatic emission bands are quite common in astronomical observations: the 3.4 \mt aliphatic C-H stretching mode and the 6.9 and 7.25~\mt bands are seen in many objects \citep[e.g.,][Jensen et al. 2018, in progress]{goto2003, goto2007, sloan2007, acke2010, sloan2014, matsuura2014, materese2017}. Aliphatics in PAHs can be present in the form of additional hydrogen attachment (superhydrogenated PAHs), which breaks the local aromaticity, or the attachment of aliphatic sidegroups, such as a methyl group. As previously stated, \citet{sloan2007} suggest that the variations in \Wpeak~may reflect different aliphatic-to-aromatic ratios: class~C objects would have an abundance of (volatile) aliphatic-bonded material, while class~A objects have a relatively low fraction of aliphatics. The premise then is that class~A environments contain PAHs that have been exposed to UV photo-processing and the relatively weak aliphatic bonds are preferentially destroyed relative to the aromatic bonds. Class~B PAH profiles would represent an intermediate photo-processing stage. In terms of direct astronomical support, \citet{goto2003} showed that the intensity of the aliphatic 3.4~\mt band relative to the aromatic 3.3~\mt band (I$_{3.4}$/I$_{3.3}$) varies with distance from the irradiating star, suggestive of a natural balance between stable aromatic material and relatively volatile aliphatic bonds. In general, astronomical observations show a I$_{3.4}$/I$_{3.3}$ varying between 0.06 and 0.20 \citep{schutte1993}. \citet{li2012} showed that these ratios indicate that $<$15\% of the carbon atoms contributing to the PAH emission bands are in aliphatic form. Taking the average per-bond integrated cross-sections for an aliphatic (A$_{3.4} = 2.7\times10^{-18}$ cm) and aromatic (A$_{3.3} = 4.0\times10^{-18}$ cm) C-H bond, we can calculate an estimated I$_{3.4}$/I$_{3.3}$ ratio for the species in our sample. We find that I$_{3.4}$/I$_{3.3}\simeq0.18\pm0.07$ for those PAHs with aliphatic sidegroups. The superhydrogenated PAHs have noticeably stronger aliphatic emission features with I$_{3.4}$/I$_{3.3}\simeq0.65\pm0.37$.

Within our sample, we identify that small ($N_{\rm c} \leq 30$) superhydrogenated PAHs are systematically redshifted from the \Wpeak~positions of small pure PAHs (cf.\ Section~\ref{sec:results_aliph}). Likewise, small PAHs with aliphatic sidegroups have \Wpeak~positions that are redshifted from small pure PAHs. These shifts are generally $\sim$0.1~\m, which is half the distance between class~A (7.6~\m), B (7.8~\m) and C ($\ge 8.0$~\m). When considering larger species, little to no shift is present: \Wpeak~measured on large superhydrogenated PAHs and PAHs with sidegroups are generally indistinguishable from their non-aliphatic counterparts. Thus again, we find that differences in size produce \Wpeak~offsets ($\sim$0.2~\m) that are larger than the redshift incurred by introducing aliphatic components ($\sim$0.1~\m).

In the literature, some tentative links between PAH class and aliphatic abundance have been presented. The aliphatic-to-aromatic flux ratio seems to increase as \Wpeak~varies from 7.9 to 8.1~\mt (\citealt{acke2010}, their Fig.~16). In addition, the classification of the 6.2~\mt PAH band, which generally mimics the redshifting nature of the 7.7~\mt band (but on a much smaller wavelength span), has been linked to aliphatics by \citet{pino2008}. These authors performed absorption spectroscopy of soot samples and found that a redshift of the 6.2~\mt band is linked to an increasing number of aliphatic-to-aromatic groups. That said, it is not yet entirely clear how aliphatics relate to class~C PAH spectra.

\paragraph{Edge structure.} The geometry of the PAHs may be relevant for determining \Wpeak. The presence of hydrogen is expected to affect the 7.7 \mt band since it originates (in part) from in-plane C-H vibrations. It is unclear if hydrogen adjacency could have an indirect effect on \Wpeak, but without question the smaller PAHs in our sample represent a much larger variety in edge structure and therefore hydrogen adjacency class compared to larger PAHs, which are predominantly symmetric/compact. Additionally, changes in edge structure are more easily induced in smaller PAHs. Thus, biases in our sample (and PAHdb itself) may be responsible for the discontinuity we find between class~A sources and large PAHs.

PAHs with bay regions are an interesting subset of species when it concerns the astronomical class~C profile. As shown by \citeauthor{peeters2017} (\citeyear{peeters2017}, their Fig.~22) the presence of bays has a strong influence on the emerging spectrum. In particular, the PAH C$_{138}$H$_{30}$ with six bay regions\footnote{Accessible through the PAHdb web interface by querying PAHdb with the unique identifier keyword (``UID''): 772, 773, 774.} has an unusual spectrum. It exhibits a single strong peak near 8.2~\m, on a modestly broad pedestal. Its appearance is similar to that of the prototypical class~C spectrum, though the class~C spectrum seems to have more broadband emission (cf.\ Fig.~\ref{fig:classes}).

\paragraph{Other factors.} It is possible that dehydrogenation affects \Wpeak~as 7.7 \mt emission is partially attributed to C-H in-plane vibrations, and we observed a broad 8~\mt band in the spectra of the fully dehydrogenated PAHs C$_{\rm 24}$, C$_{\rm 54}$, C$_{\rm 66}$, and C$_{\rm 96}$ \citep[see also][]{bauschlicher2013}. However, there is generally such a high degree of volatility in the spectra of dehydrogenated PAHs \citep{mackie2015} that it is hard to form direct links between \Wpeak~and specific astronomical PAH populations---unless a scenario akin to the grandPAH hypothesis is invoked \citep[][i.e., astronomical PAH spectra may be dominated by only a few species of very stable `prefered' PAHs]{andrews2015}.

Turning to composition, nitrogen-substituted PAHs or PANHs are a commonly considered class of heteroatomic PAHs \citep{hudgins2005,bauschlicher2008,bauschlicher2009}. We measured \Wpeak~for the PANHs present in our sample and found they all exhibited class~A (\Wpeak$\simeq$7.6 \m) peak positions, with essentially no variation in \Wpeak~as a function of ionization fraction. Other atomic substitutions, including Mg and Fe are possible. Mg-incorporated PAHs display strong emission in the 35--40 \mt range \citep{bauschlicher2009mg}, but as yet are not linked to changes in \Wpeak~(if present at all; see also \citealt{hudgins2005} who examined the 6.2~\mt band).


\subsection{\texorpdfstring{\Wpeak~}{lambda77} and implied PAH sizes}
\label{sec:discuss_size}

From an astronomical standpoint, the class~A population is expected to be associated with the largest (i.e., the most stable) PAHs. However, our results and, e.g., \citet{bauschlicher2008,bauschlicher2009}, show that class~A spectra are linked to small PAHs (\nc $\leq 60$). Motivated by this, we examine additional factors affecting the 7.7 \mt PAH complex and the implied astronomical PAH size distributions.

\begin{figure}
\centering
\includegraphics[width=1\linewidth]{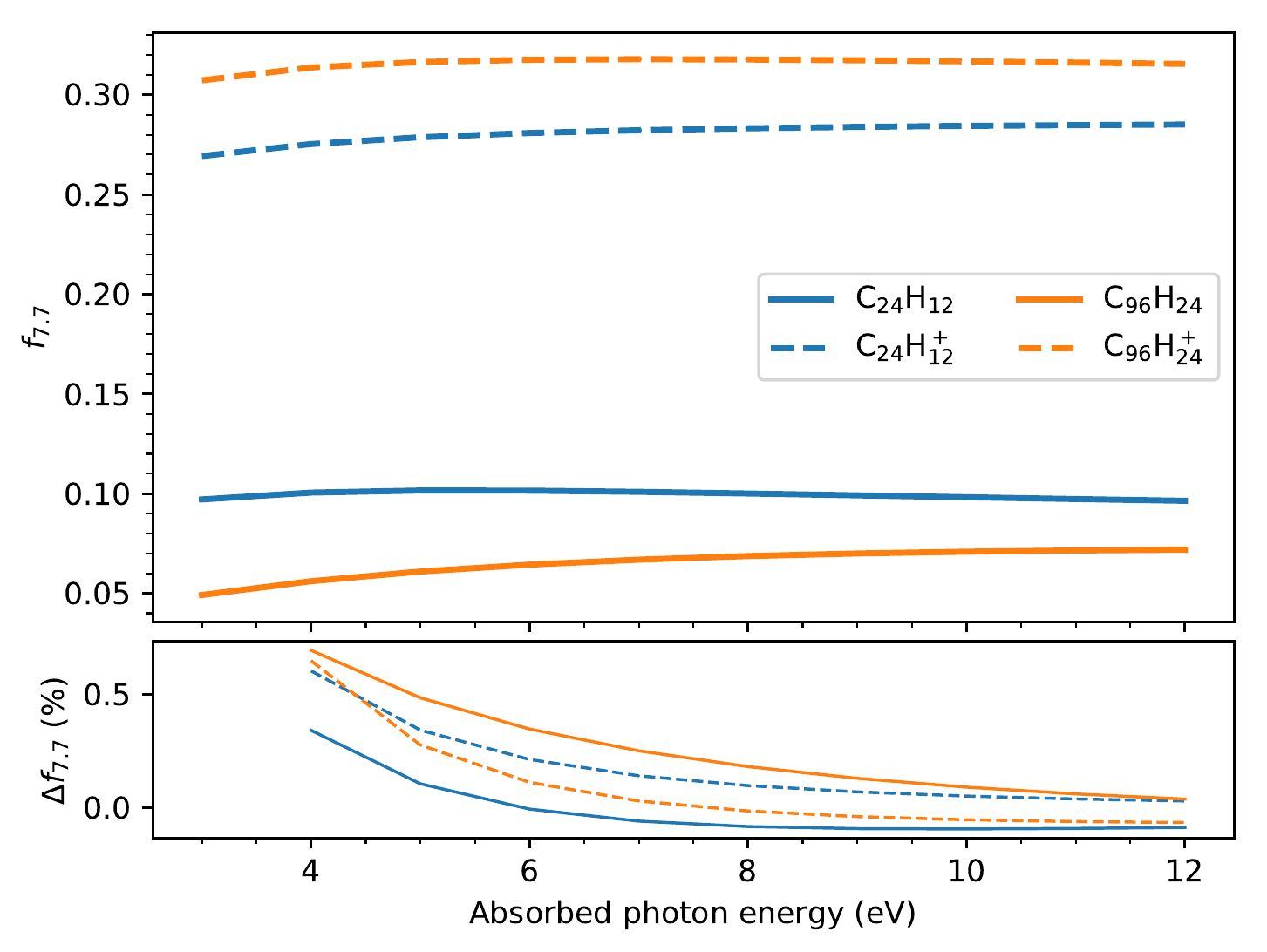}
\caption{Fractional energy emitted in the 7.7~\mt band ($f_{\rm 7.7}$) for coronene (C$_{\rm 24}$H$_{\rm 18}$) and circumcircumcoronene (C$_{\rm 96}$H$_{ \rm 24}$) after absorbing photons of different energies (3--12~eV; upper panel). The fractional energy ($f_{7.7}$) was computed by integrating the emission between 7.2 and 8.1~\m. The relative change in $f_{7.7}$ (in percent) is shown for 1~eV increments (lower panel). See Section~\ref{sec:discuss_size} for details.}
\label{fig:energy_fraction_77_band}
\end{figure}

\paragraph{PAH excitation.} First, we examine and reflect on the role PAH excitation has on \Wpeak~as a function of PAH size. We begin by reiterating the fact that size has a direct influence on the resulting PAH emission spectrum. For instance, compare coronene (C$_{24}$H$_{12}$) to circumcircumcoronene (C$_{96}$H$_{24}$). Upon absorbing a (e.g.,) 5~eV photon, coronene attains a maximum temperature of approximately 1140~K, whereas circumcircumcoronene only reaches $\sim$550~K (when using the thermal approximation; see e.g., \citealt{verstraete2001, boersma2014_amesdb}). Consequently, differences in their emission spectra are expected. Consider the relative energy emitted in the 7.7~\mt complex as a fraction of the total absorbed energy for these two PAHs (denoted by $f_{7.7}$). We present $f_{7.7}$ as a function of absorbed photon energy between 3-12~eV in Figure~\ref{fig:energy_fraction_77_band}. We compute $f_{7.7}$ by integrating the emission between two fixed wavelengths (7.2 and 8.1~\m). Neutral coronene emits approximately 10\% of its absorbed energy in the 7.7~\mt band ($f_{7.7}\simeq0.10$), whereas neutral C$_{96}$H$_{24}$ is markedly lower at 6\% ($f_{7.7}\simeq0.06$). When ionized, $f_{7.7}$ is near 30\% for both molecules, with the larger circumcircumcoronene now emitting a slightly higher fraction of its energy in the 7.7~\mt complex. For both cations, there is a very weak dependence on photon energy, generally at the sub-percent level. The takeaway is that the relative energy emitted through the 7.7~\mt band is highly dependent on ionization state, weaker on size and practically insensitive to the energy of the absorbed photon (cf.\ Fig.~\ref{fig:energy_fraction_77_band}).

\paragraph{Fitting.} We consider the relative abundances of small and large PAHs needed to match the astronomical observations. Specifically, we fit our prototypical class~A and B profiles with two (synthesized) template spectra, one representing ``small'' PAHs and the other ``large'' PAHs. We take the $\beta=0.0$ case. As such, we find good fits to the astronomical spectra when the boundary between large and small is set at $N_{\rm c}=90$. The effective PAH size within the small PAH group is $N_{\rm c}=45$ and within the large group, $N_{\rm c}=165$, which we will call the effective PAH sizes for small and large PAHs, respectively.

The best fit to the class~A observational spectrum, at $f_i=0.7$, shows that the needed ratio of small-to-large PAHs is approximately 80, i.e., almost no contribution from large PAHs. If we take this ratio, in conjunction with our effective PAH sizes, we calculate the power-law index for a size distribution that fit these data. We find that this distribution of PAH sizes can be described by a power-law with $\beta\simeq6.75$. This is clearly much steeper than the MRN distribution ($\beta=3.5$), though this is to be expected in this case, as the large molecules are not needed for a good fit. In contrast, small PAHs \emph{are} needed to produce a good fit to the astronomical class~B spectrum. We find that approximately 1.8 times as many small-to-large PAHs are needed to best fit the class~B spectrum (which we find to occur at $f_i=0.3$). In spite of the fact that the 7.8~\mt emission is generally carried by large PAHs, a greater number of small species are needed to reproduce the observational 7.7~\mt band. The associated size distribution for this ratio (1.8:1) is quite flat with $\beta\simeq0.96$.

\paragraph{Photo-destruction.} Photo-destruction could be very important in understanding the predominance of small PAHs in class~A spectra. That is, could the conversion of large PAHs into smaller PAHs produce enough small PAHs such that their summed spectrum appears as class~A? \citet{montillaud2013} modeled the evolution of PAHs using the Meudon PDR code \citep[][]{2006ApJS..164..506L, lebourlot2012} and found that much of it is driven by (multi-)photon absorption events, notably in the diffuse ISM. Nonetheless, the evolutionary timescales they determined approach, in many cases, the dynamic timescale of the system. Hence, it becomes difficult to reconcile a simple scenario in which larger PAHs are photo-destroyed into smaller PAHs in light of results from current PAH evolution models. Notably, the photo-destruction pathway for PAHs is first dehydrogenation and then subsequently losing C$_{\rm 2}$ groups (\citealt{irle2006, berne2015}; \citealt{castellanos2018} and references therein).

\subsection{Limitations}
\label{subsec:limitations}

The results and their interpretation presented here are inherently tied to the contents and breadth of the data that are contained in the spectroscopic libraries of PAHdb and the accuracy of the employed PAH emission model. As far as the content and breadth of the spectral libraries are concerned, it is very difficult to fully and quantitatively assess the influence of their (in)completeness on the derived results. Furthermore, the analysis presented here places beyond that of ``astronomical'' PAHs \citep[cf.\ ][]{bauschlicher2018} the additional constraint that both the PAH's singly charged cation and neutral spectrum needs to be available. This reduces the pool of PAH spectra under consideration from 1,877 to just under 250 and explicitly ignores any potential contribution from PAH anions to the 7.7~$\mu$m PAH complex. Although the presence of PAH anions in space has been suggested \citep[e.g.,][]{bregman2005, bauschlicher2009}, here they were disregarded in favor of a larger pool of PAH spectra to draw from. In addition, the spectroscopic libraries contain harmonic spectra and the employed PAH emission model is far from sophisticated enough to reproduce \emph{all} anharmonic aspects. 

More specifically, one of the key limitations of the spectroscopic libraries at present is that there is only a limited number of large PAHs ($N_{\rm c}>50$) with irregular edges/non-compact geometry. This results in a limited number of large PAHs with varying adjacency classes, which is particularly relevant for emission between 10--15~$\mu$m, as it is attributed to C-H out-of-plane motions of peripheral hydrogen atoms \citep[][]{bauschlicher2018, boersma2018}. However, this has potentially a big influence on the 7.7~$\mu$m PAH complex as well, as the associated emission is partially originating from C-H in-plane motions of peripheral hydrogen atoms. 

Regarding nitrogenated PAHs, superhydrogenated PAHs or PAHs with aliphatic sidegroups, the spectroscopic libraries of PAHdb only contain a few of these. As such, the limited sample hampers drawing any firm conclusion on the impact of these species on \Wpeak.

Anharmonicity adds asymmetry with extended red wings to the emission profiles, which would translate into somewhat redder~\Wpeak~positions that those reported here. However, it is difficult to predict its exact impact.

In summary, the current limitations affecting our study can largely be overcome by (1) the computation and addition of spectra from large PAHs covering all hydrogen adjacency classes and charge states to PAHdb; (2) the computation and addition of compact and irregular large PANH spectra in each charge state to PAHdb; and (3) the computation and addition of PAH spectra for large PAHs that treat anharmonicity and/or develop a PAH emission model that better treats anharmonicity. Work is currently underway dealing with each of these, as Ricca et al. (in progress) are computing the spectra for the necessary large PAHs and PANHs, and Mackie et al. (in progress) are computing anharmonic PAH spectra.

\subsection{Astronomical picture}
\label{subsec:picture}

\begin{figure*}
\centering
\includegraphics[width=0.75\linewidth]{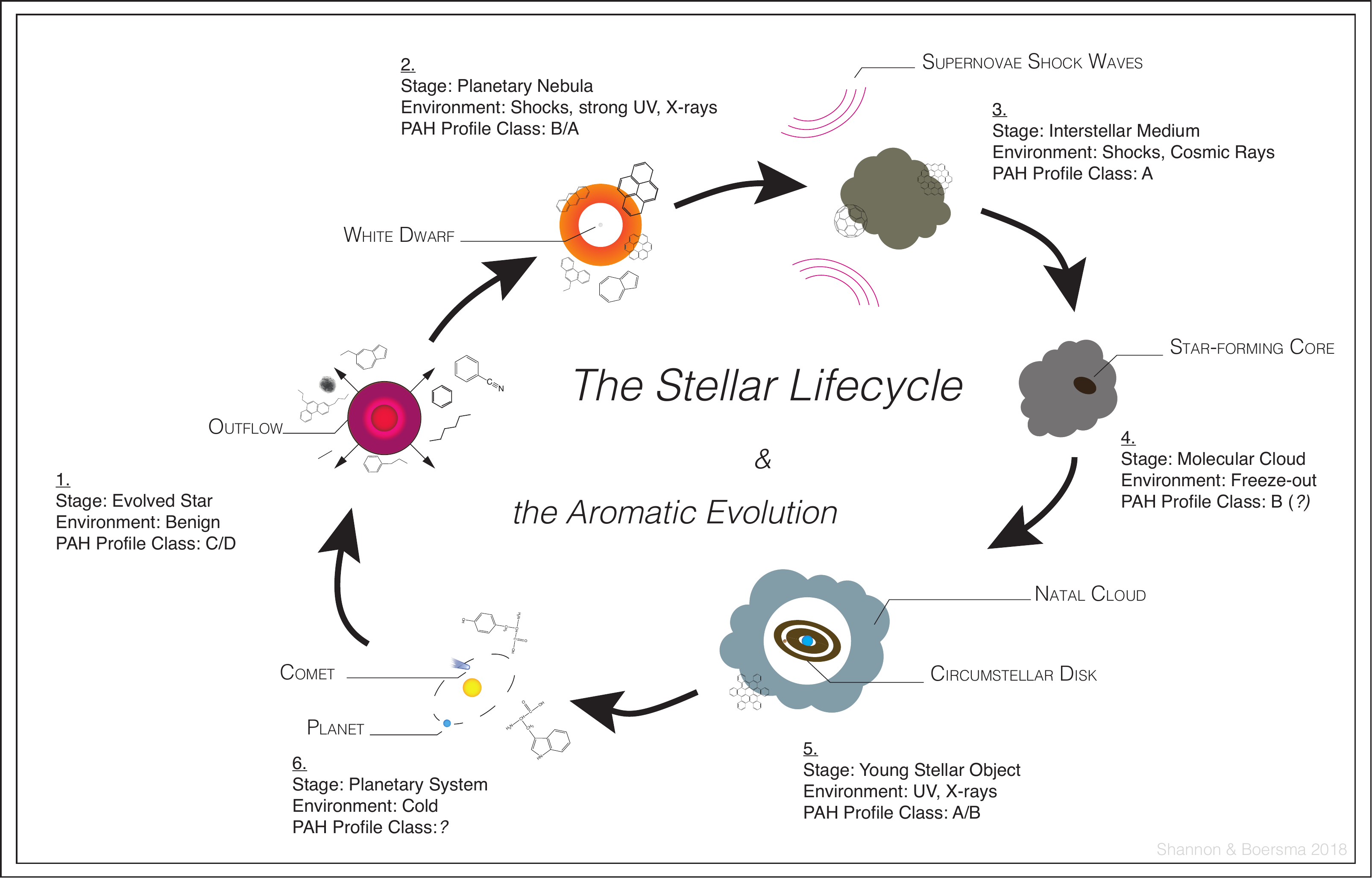}
\caption{A simple and idealized depiction of the stellar life cycle of low and intermediate mass stars, overlaid with the evolution of aromatic material, i.e., of PAH and PAH-related species. Aromatic materials are formed in the ejecta of evolved stars, processed in the interstellar medium, and incorporated into young stellar systems. The classification of the 7.7~\mt PAH complex is linked to object type: exposed interstellar environments show class~A profiles while class~B profiles are observed from circumstellar environments. See Section~\ref{subsec:picture} for a discussion.}
\label{fig:lifecycle}
\end{figure*}

Here we present a largely \emph{qualitative} sketch of the evolution of aromatic material through the life-cycle of low- and intermediate-mass stars in Fig.~\ref{fig:lifecycle}. The life-cycle of PAHs---i.e., their formation, evolution and destruction---are, predictably, dependent on the environments in which they reside. Since the distinct PAH classes defined by \citet{peeters2002}, \citet{vandiedenhoven2004} and \citet{matsuura2014} are linked to object type they must represent some manner of chemical and/or physical evolution of the PAH population when moving between object type. Thus, the following idealized picture emerges:\\

1. PAHs and other large complex molecules are thought to be formed in the ejecta of carbon-rich evolved stars through processes akin to soot chemistry and through incremental coagulation \citep[e.g.,][]{latter1991, frenklach1989, cherchneff2012}. In this ``bottom-up'' formation process, PAHs are born into a shielded environment and are thought to represent a class C population, with the general consensus being that most of these PAHs are small and contain many aliphatic chains \citep[e.g.,][]{sloan2007}. On the other hand, a ``top-down'' scenario has also been suggested, in which grains containing aromatic and aliphatic units shatter and fragment into PAHs \citep[e.g.,][]{pino2008, carpentier2012}.

2. Over time the gentle winds produced by the evolved star push the PAHs into the interstellar medium. As it transitions into a planetary nebula with a hot ($T_{\rm eff}>100,000$~K) white dwarf at its center, the system passes through an intermediate protoplanetary nebula phase. In these far more exposed environments, the PAHs are processed by energetic far-UV, extreme-UV, and X-ray photons and through shocks driven by strong jets. Here the PAH population shifts from class~C{\textrightarrow}B \citep[][]{sloan2007, sloan2014}. It is at this stage that the PAHs are likely to lose any aliphatic side chains/groups and those at the smallest end of the size distribution will be culled.

3. Eventual dissipation of nebular material puts the surviving PAHs into the interstellar medium. The exposure to energetic photons and particles whittle the PAH population down even further to only its most stable members \citep[e.g.,][]{micelotta2011}. In addition, strong interstellar shock waves from supernova explosions will affect the population, where sputtering can break down even the largest PAHs \citep[e.g.,][]{micelotta2012}. On the other hand, grain shattering and sputtering will replenish some of the population \citep[][]{jones1996}. Here the population has shifted from class~B{\textrightarrow}A and it is presumed that only the largest PAHs survive.

4. Ultimately the free-floating interstellar material will gather once more to form giant molecular clouds and in their dense star-forming cores the PAHs will freeze-out onto icy grains \citep[e.g.,][]{bouwman2011, cook2015}. During this stage processing by energetic UV and cosmic rays will process the ice-locked PAHs, adding sidegroups as well as eroding them. Though some PAHs will be liberated by the occasional interaction with a penetrating cosmic ray, most will remain in stasis. While there are astronomical observations that indeed suggest the PAHs are frozen onto ices, as of yet there is no clear signature that allows band classifications \citep[e.g.,][]{hardegreeullman2014, boogert2015}.

5. During the next phase of star formation, the forming protostar will heat up its surroundings and liberate the PAHs from their icy prisons. Despite possible alterations of the populations due to (photo-)chemical reactions with the ice \citep[e.g.,][]{bouwman2011b}, the conditions in the photo-dissociation regions associated with, for example, \hae stars are prone to drive the PAHs back to class~A; i.e., any added peripherals are quickly lost while possibly promoting H$_{\rm 2}$ formation (see, e.g., \citealt{boschman2015} and references therein). Further processing of the PAHs occurs in the photosphere of the protoplanetary disk by stellar UV photons, X-rays, and energetic particles from the protosolar wind where they transition to a class~B population \citep[see e.g.,][]{tielens2013_book}. In this environment, either the PAHs were able to retain (some of) the chemical alterations made to them in the ices or there is an on-going active chemistry (e.g., the dredge-up of material from the mid-plane to the surface of the disk).

6. Once the planetary system is formed, the PAHs will have been incorporated into the pebbles, planetesimals, and cometesimals that deliver much of the organic and volatile reservoir to emerging Earths. The PAHs themselves might even play an important role in the formation of life itself by providing a molecular framework for the construction of RNA \citep[see][]{ehrenfreund2006}. As the stellar system ages, its star will move into an evolved phase and the entire cycle repeats.\\

Our work suggests that in each of the stages outlined above it is a significant change in the PAH size distribution that is  driving the class conversion. However, it also becomes clear that the evolution of the PAH population on its route from the ejecta of evolved stars (class~C{\textrightarrow}B) into the interstellar medium (class~B{\textrightarrow}A) is different from the evolution it undergoes when moving from a circumstellar cloud (class~A) to circumstellar disk (class~B) environment.

\section{Summary and conclusions}
\label{sec:conclusions}

The observed source-to-source variations in the appearance of the astronomical 7.7~\mt PAH complex is profound. However, many years after the systematic classification by \citet{peeters2002}, and subsequent introduction of class~D by \citet{matsuura2014}, what exactly drives these variations remains largely enigmatic. Over the years, several explanations have been proposed, including changes in PAH size distributions, aliphatic content, ionization fraction or heteroatom substitution. Utilizing the data and tools made available through the NASA Ames PAH IR Spectroscopic Database
we synthesized PAH emission spectra to systematically examine the influence of different molecular parameters on \Wpeak.

We find that in our synthesized spectra \Wpeak~is most sensitive to variations in PAH size and can generally accommodate shifts of $\sim$0.4~\m. The attachment of additional hydrogen atoms or aliphatic sidegroups only produces relatively minor shifts of $\sim$0.1~\m. This is after accounting for shifts driven by size, which are $\sim$0.2~\mt in these species. Overall, the imposed PAH ionization fraction hardly affects \Wpeak, producing a maximum shift of about 0.05~\mt in pure PAHs and about 0.15~\mt in those that have an aliphatic character. The synthesized spectra of PAHs containing nitrogen all display class~A \Wpeak~profiles and appear insensitive to ionization fraction. Overall, we can accommodate class transitions from B{\textrightarrow}A, but fail with the C{\textrightarrow}B transitions. Although, PAHs with bay regions produce emission near 8.2 \mt \citep[see][]{peeters2017} that is reminiscent of a class~C emission profile (albeit slightly narrower). The synthesized PAH emission spectra show that variation in \Wpeak~can be driven, in order of maximum shift, by PAH size, aliphatic character and ionization fraction.

The astronomical picture that emerges is one where the changes the PAHs undergo on their way from the ejecta of evolved stars where they are formed (class~C{\textrightarrow}B) into the ISM (class~B{\textrightarrow}A) is different from the changes they undergo when moving from the circumstellar cloud (class~A) into the circumstellar disk (class~B) environment. The ebbing of (small) PAH sizes as they enter the interstellar medium does not work in reverse, in which interstellar PAHs are incorporated into newly-forming young stellar objects. PAH-ice chemistry may be part of the puzzle that explains the class~A{\textrightarrow}B transition of cloud and disk material, such as in the \hae stars (which show a smooth distribution of \Wpeak~and \teff).

Obviously, the results and their interpretations are inherently tied to the content and breadth of the data that is contained in the spectroscopic libraries of PAHdb and the accuracy of the employed PAH emission model. Looking forward, the addition of spectra from PAH classes that are currently underrepresented or missing (notably PAH species with a larger variation in sizes; edge structures and aliphatic content) and better treatment of anharmonicity in the PAH emission model should put our results on a stronger footing and possibly allow for a better understanding of the spectral class~B{\textrightarrow}A transition in young stellar objects. In addition, the James Webb Space Telescope will allow us to capture the full near- to mid-infrared PAH spectrum with high spatial resolution, sensitivity, and resolving power, and help better distinguish the disk and cloud components (particularly for the \hae stars).

\acknowledgements

MJS's research was supported by an appointment to the NASA Postdoctoral Program at NASA Ames Research Center, administered by the Universities Space Research Association under contract with NASA. CB is grateful for an appointment at NASA's Ames Research Center through San Jos\'e State University Research Foundation (NNX14AG80A). CB's research was supported by NASA's ADAP program (NNH16ZDA001N). This work makes use of data and tools provided by the NASA Ames PAH IR Spectroscopic Database, which is being supported through a directed Work Package at NASA Ames titled: ``Laboratory Astrophysics -- The NASA Ames PAH IR Spectroscopic Database''. We are grateful to Lou Allamandola and Jesse Bregman for the many insightful and lively discussions on the subject. Lastly, we thank the anonymous referee for her/his insightful comments and suggestions that have improved the paper.

\appendix
\renewcommand{\thefigure}{A\arabic{figure}}
\setcounter{figure}{0}

\section{Observational dataset}

Table~\ref{tab:sources} lists the astronomical data collected from the literature for a wide variety of object-types, particularly for use in the \Wpeak~vs.~\teff~plot shown in Fig.~\ref{fig:77vsteff}.

\renewcommand*{\arraystretch}{1.0}
\startlongtable
\begin{deluxetable*}{lllllll}
  \tablecaption{Data adopted from the literature.\label{tab:sources}}
  \tablehead{
    \colhead{Source}  & \colhead{Object type$^a$}  & \colhead{\Wpeak~(\m)}      & \colhead{\teff~(K)}      & \colhead{Class} & \colhead{Ref:~\Wpeak} & \colhead{Ref:~\teff}
    }
    \startdata
    BD +30 3639 & PN  & 7.842 $\pm$ 0.009 & 55000 & B1  & 1 & 16  \\
    BD +40 4124 & Non-isolated HAeBe & 7.603 $\pm$ 0.034 & 22000 & A & 1 & 5 \\
    CD -42 11721  & Non-isolated HAeBe & 7.609 $\pm$ 0.016 & 14000 $\pm$ 1000 & A & 1 & 22 \\
    CRL 2688  & Post-AGB  & 8.202 $\pm$ 0.062 & 7250  $\pm$ 400 & C & 1 & 19 \\
    G 327.3-0.5 & \HII & 7.619 $\pm$ 0.028 & 47300 & A/B1  & 1 & 8 \\
    GGD-27 ILL  & Star-forming region & 7.603 $\pm$ 0.010 & 25000 & A & 1 & 5 \\
    HD 179218 & Isolated HAeBe & 7.786 $\pm$ 0.126 & 10500 & B1/2  & 1 & 5 \\
    HD 100546 & Non-isolated HAeBe & 7.903 $\pm$ 0.136 & 10500 & B & 1 & 5 \\
    HE 2-113  & PN  & 7.913 $\pm$ 0.043 & 29000 & B2  & 1 & 9 \\
    HR 4049 & Post-AGB  & 7.869 $\pm$ 0.037 & 7500  & B2  & 1 & 5 \\
    IRAS 03260+3111 & Non-isolated HAeBe & 7.622 $\pm$ 0.027 & 12000 & A & 1 & 5 \\
    IRAS 07027-7934 & PN  & 7.921 $\pm$ 0.041 & 22000 $\pm$ 400 & B2/3  & 1 & 7 \\
    IRAS 10589-6034 & C\HII  & 7.630 $\pm$ 0.008 & 35000 & A & 1 & 8 \\
    IRAS 12063-6259 & C\HII & 7.626 $\pm$ 0.024 & 38400 & A & 1 & 8 \\
    IRAS 12073-6233 & C\HII/star-forming region & 7.626 $\pm$ 0.040 & 40200 & A & 1 & 8 \\
    IRAS 13416-6243 & Post-AGB  & 8.199 $\pm$ 0.059 & 5440 & C & 1 & 1 \\
    IRAS 15384-5348 & C\HII  & 7.618 $\pm$ 0.015 & 41000 & A & 1 & 8 \\
    IRAS 15502-5302 & C\HII  & 7.589 $\pm$ 0.027 & 36300 & A & 1 & 8 \\
    IRAS 16279-4757 & Post-AGB  & 7.633 $\pm$ 0.031 & 5700 & A & 1 & 10  \\
    IRAS 16594-4656 & Post-AGB  & 7.621 $\pm$ 0.035 & 12000$\dagger$ & A & 1 & 11  \\
    IRAS 17047-5650 & PN  & 7.830 $\pm$ 0.026 & 28200 $\pm$ 800 & B1  & 1 & 12  \\
    IRAS 17279-3350 & C\HII  & 7.622 $\pm$ 0.023 & 37300 $\pm$ 3400  & A & 1 & 4 \\
    IRAS 17347-3139 & PN  & 7.972 $\pm$ 0.033 & $\geq$ 26000 & B3  & 1 & 13  \\
    IRAS 18032-2032 & C\HII  & 7.613 $\pm$ 0.016 & 35200 $\pm$ 600 & A & 1 & 4 \\
    IRAS 18116-1646 & C\HII  & 7.630 $\pm$ 0.009 & 35500 $\pm$ 500 & A & 1 & 4 \\
    IRAS 18317-0757 & C\HII  & 7.634 $\pm$ 0.007 & 40100 & A & 1 & 8 \\
    IRAS 18434-0242 & C\HII  & 7.637 $\pm$ 0.010 & 44840 & A & 1 & 5 \\
    IRAS 18502+0051 & C\HII  & 7.623 $\pm$ 0.020 & 44500 & A & 1 & 8 \\
    IRAS 18576+0341 & LBV & 7.653 $\pm$ 0.084 & 10900 & B1  & 1 & 15  \\
    IRAS 19442+2427 & C\HII  & 7.614 $\pm$ 0.019 & 15000 $\pm$ 5000$^\dagger$ & A & 1 & 17  \\
    IRAS 21190+5140 & C\HII  & 7.612 $\pm$ 0.024 & 36000 $\pm$ 700 & A & 1 & 4 \\
    IRAS 22308+5812 & C\HII  & 7.614 $\pm$ 0.023 & 34900 $\pm$ 600 & A & 1 & 4 \\
    IRAS 23030+5958 & C\HII  & 7.625 $\pm$ 0.054 & 35700 $\pm$ 500 & A & 1 & 4 \\
    MWC 922 & Emission-line star  & 7.665 $\pm$ 0.030 & 33000 & B1  & 1 & 13  \\
    NGC 2023  & RN  & 7.609 $\pm$ 0.069 & 23000 & A & 1 & 6 \\
    NGC 7023 I  & RN  & 7.598 $\pm$ 0.022 & 17000 & A & 1 & 18  \\
    NGC 7027  & PN  & 7.814 $\pm$ 0.001 & 200000 & A & 1 & 20  \\
    Orion Bar D5 & \HII & 7.634 $\pm$ 0.007 & 37000 & A & 1 & 21  \\
    Orion Bar H2S1 & \HII & 7.627 $\pm$ 0.010 & 37000 & A & 1 & 21  \\
    S 106 (IRS4)  & YSO & 7.621 $\pm$ 0.016 & 37000 & A & 1 & 5 \\
    W 3A 02219+6125 & C\HII  & 7.626 $\pm$ 0.005 & 40100 $\pm$ 2100  & A & 1 & 4 \\
    XX-OPH  & Variable star & 7.848 $\pm$ 0.052 & 22500 $\pm$ 7500$^\dagger$ & B3  & 1 & 14 \\
    \midrule
    AFGL 2688 & Post-AGB  & 8.200 $\pm$ 0.010 & 6520 & C & 2 & 2 \\
    HD 44179  & Post-AGB  & 7.840 $\pm$ 0.060 & 9520 & B & 2 & 2 \\
    HD 100764 & Red giant & 8.150 $\pm$ 0.060 & 4850 & C & 2 & 2 \\
    HD 135344 & Herbig Ae/Be  & 8.080 $\pm$ 0.040 & 6590 & B/C & 2 & 2 \\
    HD 233517 & Red giant & 8.260 $\pm$ 0.030 & 4475 & C & 2 & 2 \\
    IRAS 13416-6243 & Post-AGB  & 8.220 $\pm$ 0.020 & 5440 & C & 2 & 2 \\
    NGC 1333 SVS 3  & RN  & 7.680 $\pm$ 0.010 & 14000 & A & 2 & 2 \\
    SU Aur  & T Tauri & 8.190 $\pm$ 0.030 & 5945 & C & 2 & 2 \\ 
    \midrule
    51 Oph  & Unknown & 8.500 $\pm$ 0.100 & 9520 & C & 3 & 3 \\
    AB Aur (HD 31293) & Herbig Ae/Be  & 7.950 $\pm$ 0.040 & 9520 & B/C & 3 & 3 \\
    Elias 3-1 (V892 Tau)  & Herbig Ae/Be  & 7.810 $\pm$ 0.040 & 10500 & B/C & 3 & 3 \\
    HD 31648 (MWC 480)  & Herbig Ae/Be  & 8.300 $\pm$ 0.100 & 8970 & C & 3 & 3 \\
    HD 32509  & Herbig Ae/Be  & 8.200 $\pm$ 0.300 & 8970 &   & 3 & 3 \\
    HD 34282  & Herbig Ae/Be  & 7.930 $\pm$ 0.020 & 9333 & B/C & 3 & 3 \\
    HD 35187  & Herbig Ae/Be  & 8.040 $\pm$ 0.010 & 8970 & B/C & 3 & 3 \\
    HD 97048  & Herbig Ae/Be  & 7.870 $\pm$ 0.030 & 9520 & B/C & 3 & 3 \\
    HD 97300  & Unknown & 7.810 $\pm$ 0.040 & 10500 & A & 3 & 3 \\
    HD 100453 & Herbig Ae/Be  & 8.080 $\pm$ 0.010 & 7390 & B/C & 3 & 3 \\
    HD 135344 & Herbig Ae/Be  & 8.050 $\pm$ 0.040 & 6590 & B/C & 3 & 3 \\
    HD 139614 & Herbig Ae/Be  & 8.040 $\pm$ 0.010 & 7850 & B/C & 3 & 3 \\
    HD 141569 & Herbig Ae/Be  & 7.960 $\pm$ 0.010 & 9520 & B/C & 3 & 3 \\
    HD 142666 & Herbig Ae/Be  & 8.060 $\pm$ 0.030 & 7580 & B/C & 3 & 3 \\
    HD 144432 & Herbig Ae/Be  & 8.390 $\pm$ 0.050 & 7390 & C & 3 & 3 \\
    HD 145718 & Herbig Ae/Be  & 8.050 $\pm$ 0.020 & 7580 & B/C & 3 & 3 \\
    HD 152404 (AK Sco)  & Herbig Ae/Be  & 8.340 $\pm$ 0.050 & 6440 & C & 3 & 3 \\
    HD 169142 & Herbig Ae/Be  & 7.990 $\pm$ 0.010 & 8200 & B/C & 3 & 3 \\
    HD 281789 & T Tauri & 7.920 $\pm$ 0.020 & 9520 & B/C & 3 & 3 \\
    SU Aur  & T Tauri & 8.180 $\pm$ 0.010 & 5945 & C & 3 & 3
    \enddata
  \tablerefs{(1) \citet{peeters2002}; (2) \citet{sloan2007}; (3) \citet{keller2008}; (4) \citealt{rudolph2006}; (5) E. Peeters (private communication); (6) \citealt{steiman-cameron1997}; (7) \citealt{surendiranath2002}; (8) \citealt{morisset2004}; (9) \citealt{demarco1998}; (10) \citealt{clayton2014}; (11) \citealt{vandesteene2003}; (12) \citealt{menzies1990}; (13) \citealt{hsia2016}; (14) \citealt{evans2012}; (15) \citealt{clark2009}; (16) \citealt{crowther2006}; (17) \citealt{ramos-larios2010}; (18) \citealt{rogers1995}; (19) \citealt{ishigaki2012}; (20) \citealt{latter2000}; (21) \citealt{simpson1986}; (22) \citealt{fernandes2007}. $^a$Object type clarifications: planetary nebula (PN), compact \HII~region (C\HII), reflection nebula (RN), luminous blue variable (LBV), isolated or non-isolated Herbig Ae/Be star (HAeBe). $^\dagger$Uncertain \teff~measurement.}
\end{deluxetable*}
\renewcommand*{\arraystretch}{1.05}

\bibliographystyle{aasjournal}

\end{document}